\documentclass[preprint,authoryear,12pt]{elsarticle}

 \usepackage{graphicx}

\usepackage{amssymb}
\usepackage{amsmath}

\usepackage[ps2pdf,%
a4paper=true,%
breaklinks=true,%
colorlinks=true,%
pdfauthor={Charles Hubaux et al.},%
pdftitle={Symplectic integration of space debris motion considering several Earth's shadowing models} %
]{hyperref}

\journal{Advances in Space Research}

\newcommand{\R}{{\mathbb R}}

\begin{document}


\thispagestyle{empty}

\vspace*{-3cm}\hspace*{-1.25cm}\begin{minipage}{14.5cm}
\includegraphics[height=3.5cm]{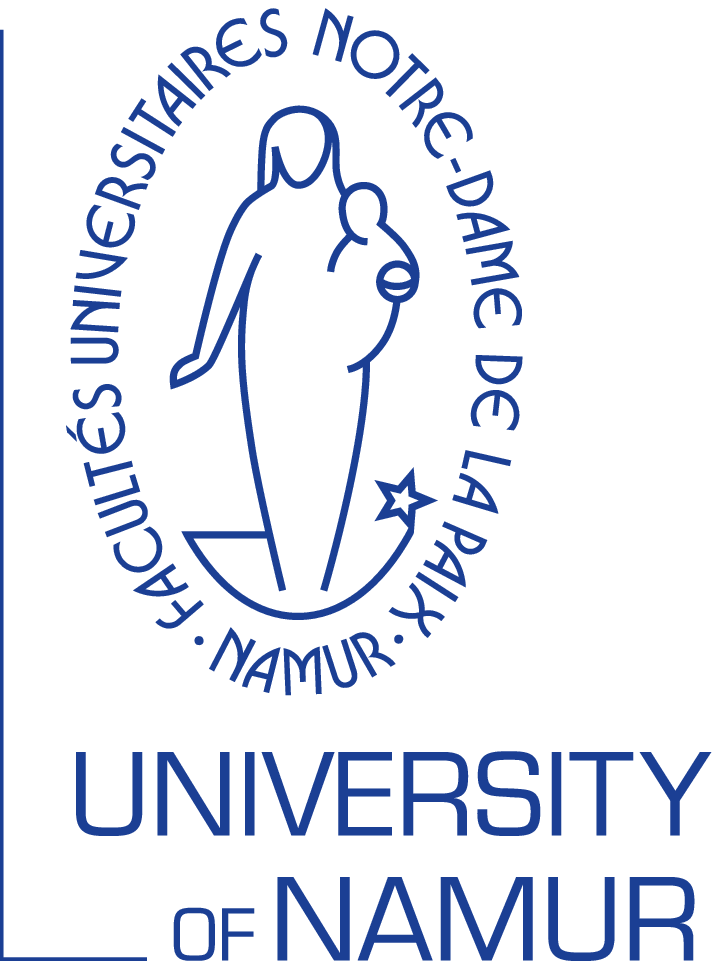}

\vspace*{1cm}
\fbox{\rule[-3cm]{0cm}{6cm}\begin{minipage}[c]{14cm}
\begin{center}
\Large Symplectic integration of space debris motion \\ considering several Earth's shadowing models\\
\mbox{}\\
by Ch. Hubaux, A. Lema\^itre, N. Delsate and T. Carletti \\
\mbox{}\\
Report naXys-01-2012 \hspace*{20mm} 19 January 2012 
\end{center}
\end{minipage}
}

\vspace{2cm}
\begin{center}
\includegraphics[height=3.5cm]{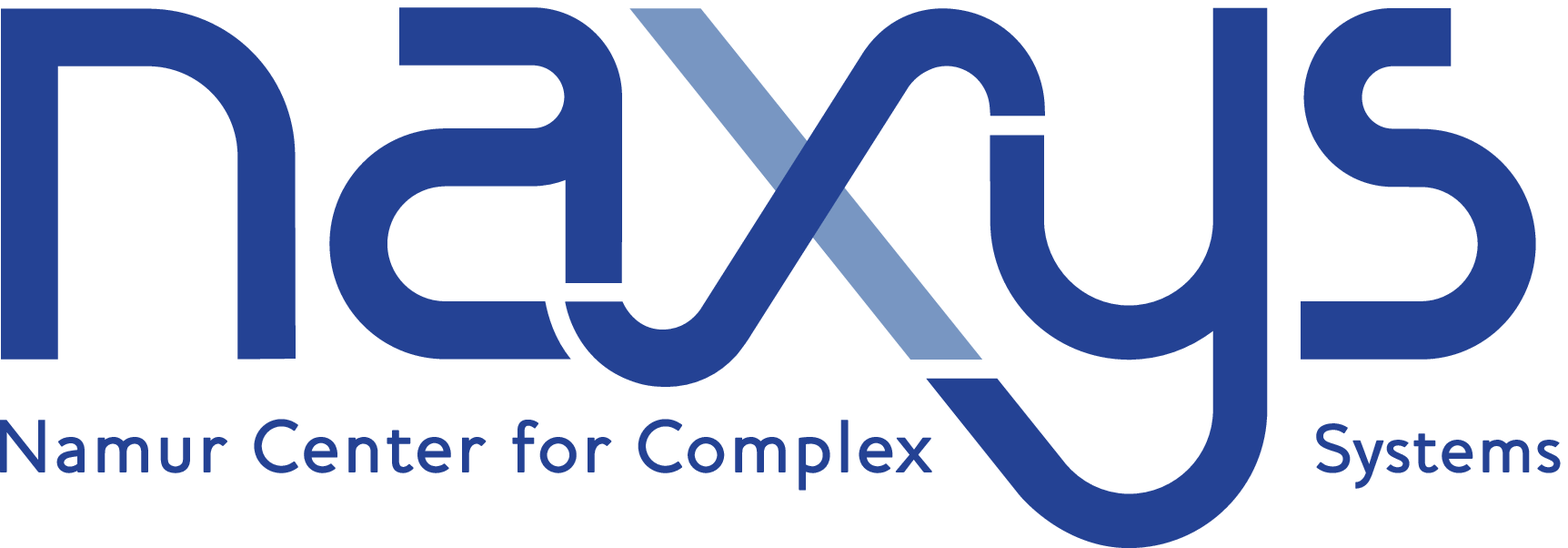}

\vspace{2cm}
{\Large \bf Namur Center for Complex Systems}

{\large
University of Namur\\
8, Rempart de la Vierge, B5000 Namur (Belgium)\\*[2ex]
{\tt http://www.naxys.be}}

\end{center}
\end{minipage}
\newpage

\setcounter{page}{1}

\begin{frontmatter}



\title{Symplectic integration of space debris motion \\ considering several Earth's shadowing models}


\author[label1]{Ch. Hubaux\corref{cor}}
\address[label1]{Namur Center for Complex Systems (NAXYS) \\ Department of Mathematics, University of Namur, \\ 8 Rempart de la Vierge, B-5000 Namur, Belgium}
\cortext[cor]{Corresponding author}
\ead{charles.hubaux@fundp.ac.be}

\author[label1]{A. Lema\^itre}

\author[label1,label2]{N. Delsate}
\address[label2]{Astronomie et Syst\`emes Dynamiques, IMCCE-CNRS UMR8028, \\ 77 Av. Denfert-Rochereau, 75014 Paris, France}

\author[label1]{T. Carletti}

\begin{abstract}

In this work, we present a symplectic integration scheme to numerically compute space debris motion. Such an integrator is particularly suitable to obtain reliable trajectories of objects lying on high orbits, especially geostationary ones. Indeed, it has already been demonstrated that such objects could stay there for hundreds of years. Our model takes into account the Earth's gravitational potential, luni-solar and planetary gravitational perturbations and direct solar radiation pressure. Based on the analysis of the energy conservation and on a comparison with a high order non-symplectic integrator, we show that our algorithm allows us to use large time steps and keep accurate results. We also propose an innovative method to model Earth's shadow crossings by means of a smooth shadow function. In the particular framework of symplectic integration, such a function needs to be included analytically in the equations of motion in order to prevent numerical drifts of the energy. For the sake of completeness, both cylindrical shadows and penumbra transitions models are considered. We show that both models are not equivalent and that big discrepancies actually appear between associated orbits, especially for high area-to-mass ratios.

\end{abstract}


\end{frontmatter}

\parindent=0.5 cm

\section{Introduction}

Several works have already shown the great importance of the direct solar radiation pressure effects on space debris motion, especially in the case of high area-to-mass ratios \citep{Liou2005,Chao2005,Anselmo2005,Chao2006,Valk2008,Lemaitre2009,Valk2009}. Moreover, some authors have pointed out that Earth's shadowing effects should not be neglected and that these phenomena could lead to significant short-periodic perturbations (see \citealt{Kozai1961}, \citealt{Ferraz1972}, \citealt{Aksnes1976} and \citealt{ValkLemaitre2008}). Recent observations \citep{Schildknecht2010} have made it possible to identify the area-to-mass ratio of $274$ uncorrelated objects. The resulting AIUB\footnote{Astronomical Institute of the University of Bern}/ESA catalogue contains a significant population of objects with area-to-mass ratios larger than $1$ m$^2$/kg and as high as $86.7$ m$^2$/kg. In this framework, perturbations due to solar radiation pressure, coupled to Earth's shadows, become essential. This paper will focus on the numerical propagation of space debris motion, considering both cylindrical shadow and penumbra transition models. 

A second aspect of the space debris dynamics that we decided to take into account in this analysis is their possible long lifetimes. If it is obvious, for low orbits,  that the efficiency of the drag forces cleans the  area in few years, it is slightly different for higher orbits, especially geostationary ones, on which the objects can stay for hundreds of years (see \citealt{Chao2005}, \citealt{Valk2008}, etc).  Up to now the numerical simulations concerning the debris  have been performed using a wide variety of numerical integrators for short timescales and the choice of a symplectic (or quasi-symplectic) integration scheme to compute the orbit of space debris has rarely been done (see \citealt{Breiter2005} for an application of a fourth order symplectic integrator of the Wisdom-Holman type). Nevertheless, it is well known that these algorithms enable the use of large time steps, show excellent energy preservation properties on long time scales and are less time-consuming than non-symplectic schemes. 

The introduction of Earth's shadow in this context is a challenging problem. Indeed, the actual direct solar radiation pressure reaching the satellite has to be computed with a sufficiently smooth function and included directly in the equations of motion in order not to introduce numerical errors in the symplectic scheme.  

The idea of the introduction of a continuous shadow function equal to one in direct sunlight and zero otherwise has been first proposed in \citet{Ferraz1964} and \citet{Ferraz1965}. This function depends on the angle formed by the geocentric Cartesian position of space debris and the dark pole of the Earth's terminator. Another approximation of this shadow function has been proposed later on in \citet{Lala1969}. Then, a way of computing shadow boundaries has been proposed in \citet{Escobal1976}. It has to be noted that, with such approximations, only cylindrical shaped Earth's shadows could be modeled. Hence, these solutions have been improved further in order to take into account penumbra transitions. For example, umbra and penumbra cone boundaries have been computed in \citet{Escobal1976}. Other detailed studies exist. For example, penumbra transitions and several physical processes in the atmosphere have been added in \citet{Vok1993} to provide a realistic shadow crossing model. This theory being really time consuming, an approximate version has been presented in \citet{Vok1994a}. Penumbra phenomena induced by the solar radiation pressure from the Earth-reflected sunlight have also been studied in \citet{Vok1994b}, leading to the conclusion that these effects were much less important than the direct solar radiation pressure. A further numerical investigation \citep{Vok1996} has also shown that the oblateness of the Earth did not bring significant differences, compared to a spherical Earth. In light of this, the oblateness of the Earth has not been included in our shadow models. Eventually, we can find another geometrical model of the penumbra transition in \citet{Montenbruck2005}, where the computation of the degree of occultation of the Sun by the Earth is performed by means of apparent radii in geocentric Cartesian coordinates.

In this paper, we present an innovative theory to model both cylindrical and conical shadows (umbra-penumbra transitions) crossings. Our model being totally included in the equations of motion of the space debris, this is well adapted to our symplectic integration scheme. Besides, our formulation has the advantage to build a smooth shadow function. Hence, it can also be included in the variational equations whose solutions are required by a variety of chaos indicators. It should be noted that it cannot be achieved with the formulation from \citet{Montenbruck2005}.

The paper is organized as follows. In Sec. \ref{secModel}, we describe the forces considered in our Hamiltonian model. Sec. \ref{secInt} is devoted to the presentation of the symplectic integrator (Sec. \ref{SecLaskar}), its implementation (Sec. \ref{SecImpl}) and shows a comparison with an efficient non-symplectic integrator (Sec. \ref{SecSimu}). Our shadow modelling theory is then developed in Sec. \ref{secShad}. First (Sec. \ref{SecCyl}), we present the basis of the model for a cylindrical-shaped Earth's shadow. Second, details are given about the adaptation of this theory in the case of penumbra transitions in Sec. \ref{SecPen}. For the sake of clarity, full mathematical developments have been moved to different appendices. Eventually, we compare our penumbra transition model to the one proposed in \citet{Montenbruck2005} in the case of a typical shadow crossing occuring on a geostationary orbit (Sec. \ref{secComp}). We also show the difference between the evolution of the orbital elements of a space debris crossing either cylindrical or conical Earth's shadows.

\section{Model}
\label{secModel}

Let us consider the following autonomous Hamiltonian function
\begin{equation*}
  \begin{array}{rcl}
    \mathcal H({\mathbf v},\Lambda,{\mathbf r},\theta) & = & \mathcal H_{\text{kepl}}({\mathbf v},{\mathbf r})+\mathcal H_{\text{rot}}(\Lambda) + \mathcal H_{\text{geopot}}({\mathbf r},\theta) \\
    &  & + \mathcal H_{\text{3body}}({\mathbf r}) + \mathcal H_{\text{srp}}({\mathbf r})
    \end{array}
\end{equation*}

where ${\mathbf r}:=(x,y,z)$ and ${\mathbf v}$ are respectively the Cartesian geocentric coordinates and velocities of the satellite, $\theta$ is the Greenwich sidereal time and $\Lambda$ is its associated momentum. Our model takes into account the attraction of the Earth as a point mass central body ($\mathcal H_{\text{kepl}}$), the rotation of the Earth around itself ($\mathcal H_{\text{rot}}$) and the perturbations due to the Earth gravity field ($\mathcal H_{\text{geopot}}$), third bodies (mainly the Sun and the Moon) ($\mathcal H_{\text{3body}}$) and the solar radiation pressure ($\mathcal H_{\text{srp}}$).

The attraction of the Earth as a central body is accounted for as 
\begin{equation*}
 \mathcal H_{\text{kepl}}=\dfrac{v^2}{2}-\dfrac{\mu}{r}
\end{equation*}
where $r:=\|{\mathbf r}\|$, $v:=\|{\mathbf v}\|$ and $\mu=\mathcal G M_\oplus$ is the standard gravitational coefficient. 

The rotation of the Earth around its axis of smallest inertia is modeled as
\begin{equation*}
  H_{\text{rot}}=\dot\theta\Lambda.
\end{equation*}

The complete Earth's potential expressed in the frame rotating around the Earth's axis of smallest inertia and with the same angular speed can be written as
\begin{equation*}
  \begin{array}{rcl}
    U_{\text{geopot}}(r,\lambda,\phi) & = & -\dfrac \mu r\sum\limits_{n=0}^\infty \sum\limits_{m=0}^n \left( \dfrac{R_\oplus}{r} \right)^n \mathcal P_{nm}(\sin \phi) \\
    & & \times (C_{nm} \cos m\lambda + S_{nm}\sin m\lambda)
  \end{array}
\end{equation*}

where $(r,\lambda,\phi)$ are the geocentric spherical coordinates of the satellite, $R_\oplus$ represents the equatorial radius of the Earth, $\mathcal P_{nm}$ are Legendre functions and $C_{nm}$ and $S_{nm}$ are the spherical harmonics coefficients.

Thanks to formulas presented in \citet{Cunningham1970} and later in \citet{Montenbruck2005}, we are able to write $U_{\text{geopot}}$ as a function of ${\mathbf r}$

$$
 U_{\text{geopot}}({\mathbf r})=-\dfrac{\mu}{R_\oplus}\sum\limits_{n=0}^\infty\sum\limits_{m=0}^n C_{nm}V_{nm}({\mathbf r})+S_{nm}W_{nm}({\mathbf r}).
$$
Let us remark that $V_{nm}$ and $W_{nm}$ depend directly on Cartesian coordinates and are defined recursively. Also note that $U_{\text{geopot}}$ is still written in the same rotating frame as the one described above. The following simple change of variable 
\begin{eqnarray*}
  x & \mapsto & x \cos \theta + y \sin \theta \\
  y & \mapsto & -x \sin \theta + y \cos \theta 
\end{eqnarray*}
lets us express $U_{\text{geopot}}$ in the fixed inertial geocentric frame.

Also note that the zero degree is already present in $\mathcal H_{\text{kepl}}$. Moreover, given that the center of mass coincides with the origin of the reference frame $C_{10}=C_{11}=S_{10}=S_{11}=0$. Hence, the sum in $U_{\text{geopot}}$ should start at $n=2$. It follows that, 
$$
 \mathcal H_{\text{geopot}}({\mathbf r})=-\dfrac{\mu}{R_\oplus}\sum\limits_{n=2}^\infty\sum\limits_{m=0}^n C_{nm}V_{nm}({\mathbf r},\theta)+S_{nm}W_{nm}({\mathbf r},\theta).
$$
The \emph{EGM96} model \citep{Lemoine1970} of degree and order $360$ is used as Earth gravity model.

Perturbations due to third bodies (the Sun, the Moon and planets of the solar system) are introduced in the Hamiltonian function as 
$$
 \mathcal H_{\text{3body}}=-\sum\limits_i \mu_i\left( \dfrac{1}{\|{\mathbf r}-{\mathbf r_i}\|}-\dfrac{{\mathbf r}\cdot{\mathbf r_i}}{\|{\mathbf r_i}\|^3} \right)
$$
where $\mu_i=\mathcal G M_i$ and ${\mathbf r_i}$ is the geocentric Cartesian coordinates of any third body of mass $M_i$. Position and velocity of the latter are computed by means of Jet Propulsion Laboratory (JPL) \emph{DE405} planetary and lunar ephemeris \citep{Standish1998}. This external contribution introduces a time dependence in the Hamiltonian, formulated as a quasi periodic function. For simpler models of the solar orbit, limited to three frequencies for example, we could have introduced three new variables (linear functions of time) and associated conjugated momenta (as it has been done for the rotation of the Earth). However it would have considerably increased the number of the differential equations to deal with, and the model would stay approximate. This is the reason for which we have chosen to work on with a quasi-symplectic formulation in this context. 

Eventually, our model also includes the direct solar radiation pressure. Three standard physical models are used: the absorption, the reflection and the diffusion. Hence, the force induced by the radiation pressure is obtained by adding together the elementary forces accounting for each effect. The model assumes that space debris are spherical objects and that there is no radiation from the surface of the Earth. Detailed information about the model can be found in \citet{Milani2009}. It leads to the following potential:
$$
 \mathcal H_{\text{srp}}=-\dfrac{1}{\|{\mathbf r}-{\mathbf r}_\odot\|}P_r\dfrac{A}{m}a^2_\odot
$$
where ${\mathbf r}_\odot$ is the geocentric Cartesian position of the Sun, $P_r=4.56\times 10^{-6}\, N/m^2$ is the radiation pressure for an object located at a distance of 1 AU from the Sun, $A/m$ is the area-to-mass ratio of the space debris and $a_\odot$ is equal to the mean distance between the Sun and the Earth (i.e. $a_\odot=1$ AU). The construction of this potential is also explained in e.g. \citet{Montenbruck2005}. Let us remark, that, making some assumptions, $\mathcal H_{\text{srp}}$ is written under its conservative form. Moreover, the shadow of the Sun by the Earth is not yet considered.

\section{Numerical integration scheme}
\label{secInt}

\subsection{Symplectic integrator}
\label{SecLaskar}

Let ${\mathcal H}(\bf p,\bf q)=\mathcal A+\varepsilon \mathcal B$ be an autonomous Hamiltonian with $N$ degrees of freedom where $\bf p,\bf q\in \R^N$ are respectively momenta and variable vectors. The Hamiltonian vector field can be written as 
$$ 
\dot{\mathbf x} = \{\mathcal H,{\mathbf x}\}=L_{\mathcal H}{\mathbf x}=\sum_{j=1}^N \frac{\partial \mathcal H}{\partial {\mathbf p}_j} \frac{\partial {\mathbf x}}{\partial {\mathbf q}_j} - \frac{\partial \mathcal H}{\partial {\mathbf q}_j} \frac{\partial {\mathbf x}}{\partial {\mathbf p}_j} 
$$
where
$$
 {\bf x}=\left( \begin{array}{c} {\bf p} \\ {\bf q} \end{array} \right) \in \R^{2N}.
$$
The solution of this differential equation is given by
\begin{equation}
  {\mathbf x}(t)=e^{tL_{\mathcal H}}{\mathbf x}(t_0). \label{solDE} 
\end{equation}

The Campbell-Baker-Hausdorff theorem \citep{Bourbaki1972} ensures that we can find a general integrator with $n$ steps of the form
$$
S_n(\tau)=e^{c_1\tau L_{\mathcal A}}e^{d_1\tau L_{\varepsilon \mathcal B}} \cdots e^{c_n\tau L_{\mathcal A}}e^{d_n\tau L_{\varepsilon \mathcal B}}=e^{\tau L_{\mathcal K}}
$$
where $\tau$ is the time span covered by the integration process.
Coefficients $c_i$ and $d_i$ are to be chosen carefully, so that the order of the integrator is improved on.
In this case, integrating $\mathcal H$ at order $m$ means that we exactly evaluate $e^{\tau L_{\mathcal K}}$
where
$$ 
\mathcal K=\mathcal A+\varepsilon \mathcal B+\mathcal O(\tau^m).
$$

In \citet{Laskar2001}, four classes of symmetric symplectic integrators are presented. 
$$
\begin{array}{rcl}
  \text{SABA}_{2n}(\tau) & = & e^{c_1\tau L_{\mathcal A}}e^{d_1\tau L_{\varepsilon \mathcal B}}... \\
                        &  & \times \, e^{c_n\tau L_{\mathcal A}}e^{d_n\tau L_{\varepsilon \mathcal B}}e^{c_{n+1}\tau L_{\mathcal A}}e^{d_n\tau L_{\varepsilon \mathcal B}}e^{c_n\tau L_{\mathcal A}} \\
                        &  & \times \, ...e^{d_1\tau L_{\varepsilon \mathcal B}}e^{c_1\tau L_{\mathcal A}} \\ 
 \text{SABA}_{2n+1}(\tau) & = & e^{c_1\tau L_{\mathcal A}}e^{d_1\tau L_{\varepsilon \mathcal B}}... \\
                         &  & \times \, e^{c_{n+1}\tau L_{\mathcal A}}e^{d_{n+1}\tau L_{\varepsilon \mathcal B}}e^{c_{n+1}\tau L_{\mathcal A}} \\
                         &  & \times \, ...e^{d_1\tau L_{\varepsilon \mathcal B}}e^{c_1\tau L_{\mathcal A}} \\
 \text{SBAB}_{2n}(\tau) & = & e^{d_1\tau L_{\varepsilon \mathcal B}}e^{c_2\tau L_{\mathcal A}}e^{d_2\tau L_{\varepsilon \mathcal B}}... \\ 
                        &   & \times \, e^{c_{n+1}\tau L_{\mathcal A}}e^{d_{n+1}\tau L_{\varepsilon \mathcal B}}e^{c_{n+1}\tau L_{\mathcal A}} \\
                        &  & \times \, ...e^{d_2\tau L_{\varepsilon \mathcal B}}e^{c_2\tau L_{\mathcal A}}e^{d_1\tau L_{\varepsilon \mathcal B}} \\
 \text{SBAB}_{2n+1}(\tau) & = & e^{d_1\tau L_{\varepsilon \mathcal B}}e^{c_2\tau L_{\mathcal A}}... \\
                         &  & \times \, e^{d_{n+1}\tau L_{\varepsilon \mathcal B}}e^{c_{n+2}\tau L_{\mathcal A}}e^{d_{n+1}\tau L_{\varepsilon \mathcal B}} \\
                         &  & \times \, ...e^{c_2\tau L_{\mathcal A}}e^{d_1\tau L_{\varepsilon \mathcal B}} 

\end{array}
$$

It is shown that, if $\varepsilon$ is small enough, one can find a suitable set of coefficients $\{c_i,d_i\}$ such that
\begin{equation}
\mathcal K=\mathcal A+\varepsilon \mathcal B+\mathcal O(\tau^{2n}\varepsilon+\tau^2\varepsilon^2). \label{eqErrorLaskar}
\end{equation}
As a matter of fact, the order of the error does not only depend on the time step anymore but also on $\varepsilon$. It yields excellent energy preservation properties and enables us to use bigger time steps than with any other non-symplectic integration scheme.

\subsection{Implementation}
\label{SecImpl}

As explained in Section \ref{SecLaskar}, the Hamiltonian function $\mathcal H$ has to be split into two integrable parts in order to be expressed as the perturbation of an integrable one. Moreover, the $\mathcal B$-part must always be smaller than the $\mathcal A$-part (i.e. $\varepsilon:=|\mathcal B|/|\mathcal A|$ is small enough). A convenient possibility is the following one:
\begin{equation*}
 \mathcal H({\mathbf v},\Lambda,{\mathbf r},\theta) =  \mathcal A({\mathbf v},{\mathbf r},\Lambda) + \mathcal B({\mathbf r},\theta)
\end{equation*}
where
\begin{eqnarray*}
 \mathcal A({\mathbf v},{\mathbf r},\Lambda) & = & \mathcal H_{\text{kepl}}({\mathbf v},{\mathbf r})+\mathcal H_{\text{rot}}(\Lambda) \\
 \mathcal B({\mathbf r},\theta) & = & \mathcal H_{\text{geopot}}({\mathbf r},\theta) + \mathcal H_{\text{3body}}({\mathbf r}) + \mathcal H_{\text{srp}}({\mathbf r})
\end{eqnarray*}
In this case, each perturbation is put in the $\mathcal B$-part and is of lower order than the energy associated to the Keplerian problem (i.e. the $\mathcal A$-part).

Each time that the operator $e^{c_i\tau L_A}$ is applied on the state vector $({\mathbf v},\theta,{\mathbf r},\Lambda)^T(t_0)$, the resulting vector corresponds to the solution of the Keplerian problem at time epoch $t_0+c_i\tau$. This computation is performed analytically in order to reduce numerical cost and keep a high accuracy. Moreover, the sidereal time is increased linearly
$$
 \theta(t_0+c_i\tau)=\theta(t_0)+c_i\tau \dot\theta 
$$
where $\dot\theta$ is assumed to be constant.

On the other hand, the solution of $e^{d_i\tau L_{\varepsilon \mathcal B}}({\mathbf v},\theta,{\mathbf r},\Lambda)^T(t_0)$ is computed numerically as the solution of the set of differential equations
$$
\left\{
\begin{array}{rcccl}
\dot {\bf v}  & = & -d_1\varepsilon {\bf \nabla}_{\bf r} \mathcal B({\mathbf r},\theta) & &  \\
\dot {\Lambda} & = & -d_1\varepsilon {\bf \nabla}_{\theta} \mathcal B({\mathbf r},\theta) & & \\
\dot {\bf r}  & = &  d_1\varepsilon {\bf \nabla}_{\bf v} \mathcal B({\mathbf r},\theta) & = & {\bf 0}  \\
\dot {\theta} & = &  d_1\varepsilon {\bf \nabla}_{\Lambda} \mathcal B({\mathbf r},\theta) & = & 0
\end{array}
\right. .
$$
It follows that 	
\begin{eqnarray*}
\hspace*{-1cm}\left( \begin{array}{c}
{\mathbf v} \\ \Lambda \\ {\mathbf r} \\ \theta
\end{array} \right)(t_0+d_i\tau)
 & = & 
e^{d_1\tau L_{\varepsilon B}}
\left( \begin{array}{c}
{\mathbf v} \\ \Lambda \\ {\mathbf r} \\ \theta
\end{array} \right)(t_0) \\
 & = & 
\left(
\begin{array}{c}
\mathbf{v}(t_0)-d_1 \varepsilon \tau \mathbf{\nabla}_\mathbf{r} \mathcal B ({\mathbf r}(t_0),\theta(t_0))  \\
\Lambda(t_0)-d_1 \varepsilon \tau \mathbf{\nabla}_\mathbf{\theta} \mathcal B ({\mathbf r}(t_0),\theta(t_0))  \\
\mathbf{r}(t_0) \\
\theta(t_0)
\end{array}
\right) 
\end{eqnarray*}

\subsection{Numerical simulation}
\label{SecSimu}

\begin{figure}
\begin{center}
\includegraphics*[width=9cm]{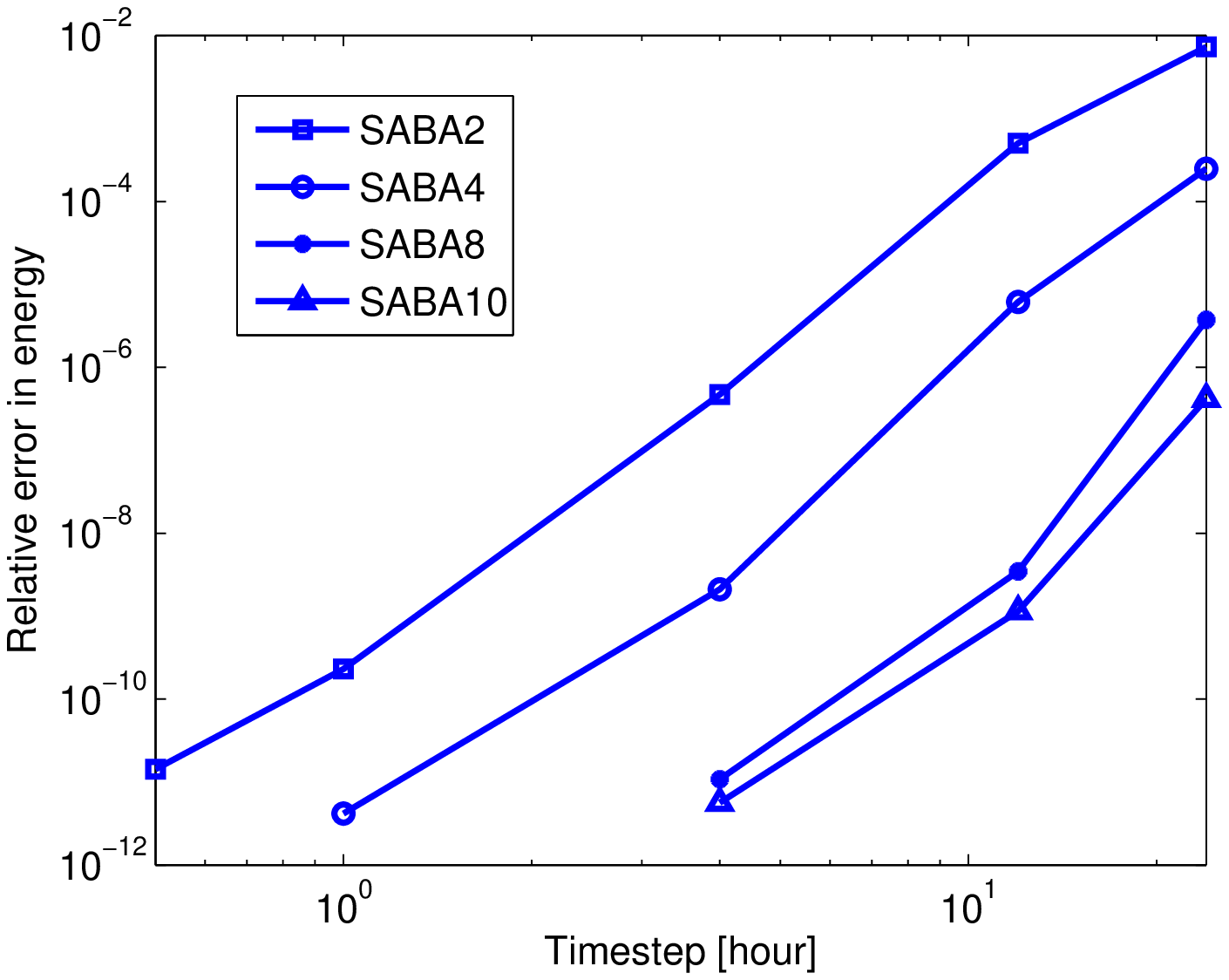} \\
\includegraphics*[width=9cm]{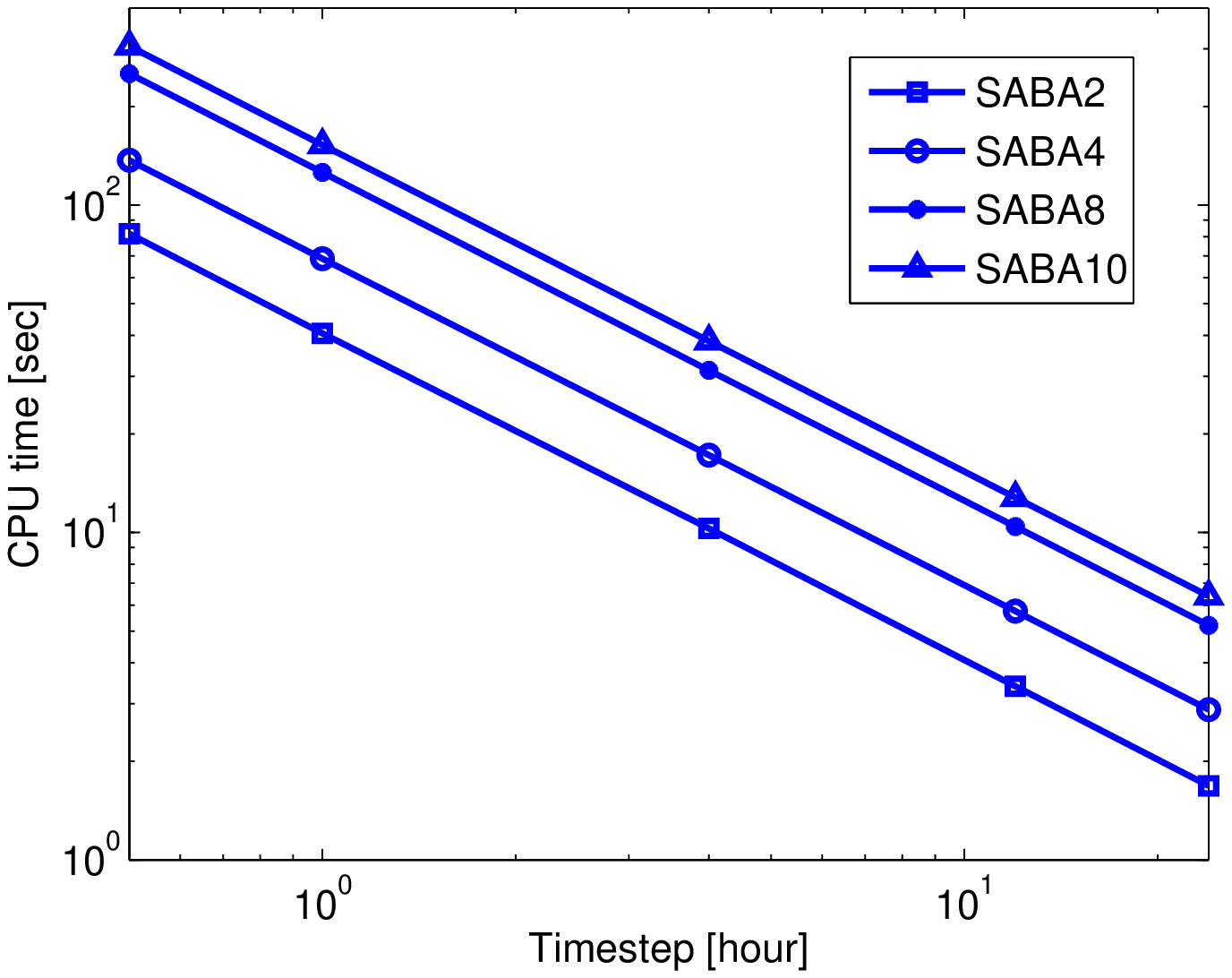}
\end{center}
\caption{Maximum relative errors in energy (top panel) and CPU times (bottom panel) for different integrator orders, as a function of the time step. Initial conditions are $a=42164.140$ km, $e=0.1$, $i=0.1$ rad, $\Omega=\omega=M=0$ rad and the initial Julian date is $2455194.5$ days. The model includes the geopotential up to degree and order $4$. The integration has been performed on a time span of $500$ years.}
\label{figRelErrEnergyCpuGeopot}
\end{figure}

This section aims to prove the numerical efficiency of our integrator. First, we show in Fig. \ref{figRelErrEnergyCpuGeopot} the maximum relative error in energy and CPU time required by our symplectic scheme to propagate the orbit of a space debris on a long time span ($500$ years), depending on several time steps and integrator orders. Initial conditions are $a=42164.140$ km, $e=0.1$, $i=0.1$ rad, $\Omega=\omega=M=0$ rad and the initial Julian date is $2455194.5$ days. The model includes the geopotential up to degree and order $4$. Time steps for these simulations have been set to $24$, $12$, $4$, $1$ and $1/2$ hours. Several observations can be made from Fig. \ref{figRelErrEnergyCpuGeopot}. On the one hand, the system being fully symplectic in this case, the energy is accurately preserved even with large time steps. Let us point out that we could not go below the accuracy treshold of $10^{-12}$. On the other hand, CPU times are really small, keeping in mind that the orbit has been propagated over $500$ years. Computations have been performed on a E5440 Intel Xeon CPU ($2.83$ GHz) with $6144$ KB cache size. 

\begin{figure}
\begin{center}
\includegraphics*[width=9cm]{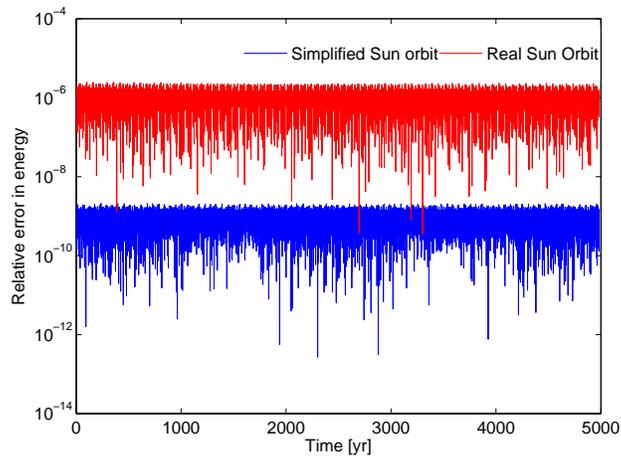}
\end{center}
\caption{Comparison of the relative errors in energy as a function of time. Both real and simplified (i.e. $e_\odot=0$ and $i_\odot=0$ rad) orbits of the Sun are considered. Initial conditions are $a=42164.140$ km, $e=0.1$, $i=0.1$ radians, $\Omega=\omega=M=0$ radians and the initial Julian date is $698382.5$ days. The model includes the geopotential up to degree and order $4$ and the graviational perturbation of the Sun. The simulation has been run with the $SABA_4$ integrator.}
\label{figRelErrEnergySun}
\end{figure}

The introduction of the gravitational perturbation of the Sun does not prevent the energy to be preserved on long time scales. However, the quasi-symplecticity of our integrator means that relative variations of the energy are about $10^{-6}$ for the third body and solar radiation pressure perturbations, which is the amplitude of the quasi periodic solar motion. In light of this, we compare the relative energy error associated to two different orbits. The model still includes the geopotential up to degree and order $4$ and we add the solar graviational perturbation. The difference appears in the computation of the Cartesian position of the Sun. On the one hand, we use the JPL ephemeris and, on the other hand, we consider a Keplerian solar motion with eccentricity and inclination equal to zero (part of the ephemeris module of the {\sc NIMASTEP} software described below). This lets us reduce the quasi periodic motion of the Sun. Results are illustrated in Fig. \ref{figRelErrEnergySun}, where the $SABA_4$ integrator has been used. Obviously, a circular and coplanar solar orbit reduces the relative error in energy to order $10^{-9}$. Fig. \ref{figRelErrEnergySun} also emphasizes the excellent energy preservation over the years, even when JPL ephemeris are used. 

It has to be noted that the introduction of the JPL ephemeris greatly increases the computation cost. As a point of comparison, the CPU time needed for this simulation after $500$ years with a time step of $4$ hours and the $\text{SABA}_4$ integrator is $22$ seconds. It corresponds roughly to twice the amount of CPU time corresponding to the same propagation without the computation of the Sun's position. 

Eventually, we compare our integration scheme to the {\sc NIMASTEP} software \citep{Delsate2011}. {\sc NIMASTEP} (Numerical Integration of the Motion of Artificial Satellites orbiting a TElluric Planet) is an extensive tool that allows to integrate numerically the osculating motion of an arbitrary object (natural or artificial satellite, space debris...) orbiting a central body ((dwarf-)~planets or asteroids of the Solar System) taking into account a large number of forces, integrators and options. This software has been successfully validated and compared to other external softwares. In the following comparisons, {\sc NIMASTEP} is used with the Adams-Bashforth-Moulton integrator of order ten \citep{Hairer1993}, hereafter referred to as ABM10. 
Starting from the same initial conditions, both integrators have been used to numerically propagate the orbit. For the sake of completeness, all kind of perturbations have been considered. Hence, we take into account the geopotential up to degree and order $4$, luni-solar perturbations and the solar radiation pressure with $A/m=0.01$ m$^2$/kg. In this case, the fourth order $\text{SABA}$ integrator has been used with time steps equal to $4$ hours. The CPU time required to propagate the initial conditions to $190$ years with this method is $11.54$ s. Table \ref{tableCpuNimastep} lists each CPU time required by {\sc NIMASTEP} for different time steps. In order to get an idea of the number of steps per revolution (about one day for these initial conditions), time steps are given in terms of fractions of one day and in seconds. This table let us point out that, even if {\sc NIMASTEP} performs the numerical integration quickly, the fact that our symplectic scheme can use large (ten times bigger) time steps make it obviously a faster algorithm.

\begin{table}  
  \begin{center}
    \begin{tabular}{ccc}
      \hline
      Step size [day] & Step size [s] & CPU time [s] \\
      \hline
      $1/200$ & $432$ & $142.01$ \\
      $1/100$ & $864$ & $79.31$ \\
      $1/86$ & $1004.65$ & $70.73$ \\            
      $1/75$ & $1152$ & $66.83$ \\
      \hline
    \end{tabular}
  \end{center}
  \caption{CPU times required by {\sc NIMASTEP} (with ABM10)  with respect to time steps.}
  \label{tableCpuNimastep}
\end{table}

\begin{figure*}
\begin{center}
\hspace*{-1cm}\includegraphics*[width=15cm]{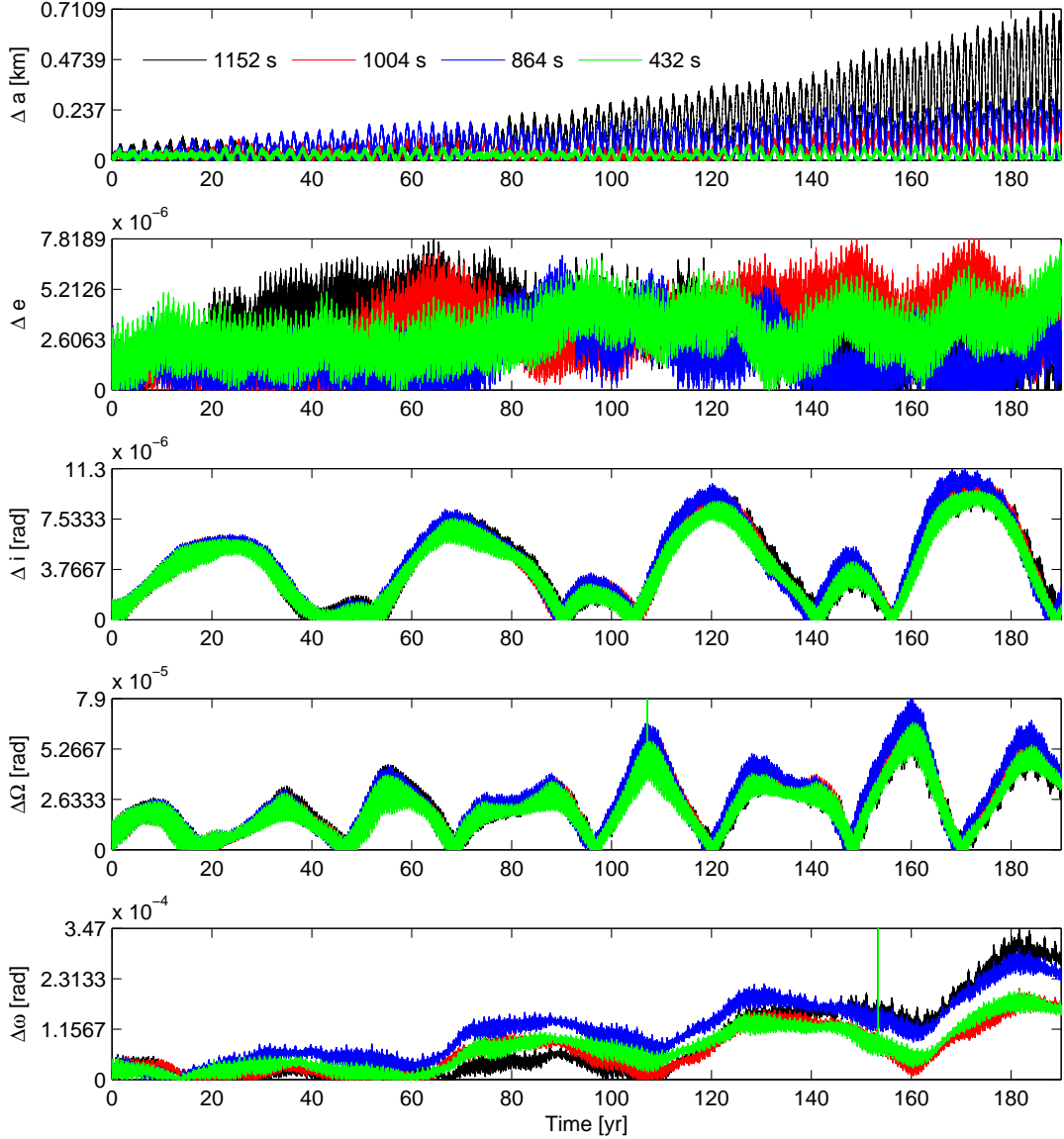}
\end{center}
\caption{Comparison of the absolute difference between each Keplerian elements of the orbit obtained by our scheme and by the {\sc NIMASTEP} software (with ABM10). Several time steps have been used with {\sc NIMASTEP}: $432$ s (green), $864$ s (red), $1004.65$ s (blue) and $1152$ s (black). Initial conditions are $a=42164.140$ km, $e=0.1$ , $i=0.1$ rad, $\Omega=\omega=M=0$ rad. The model includes the geopotential up to degree and order 4, solar radiation pression (with $A/m=0.01$ m$^2$/kg) and luni-solar perturbations. The initial Julian date is $2455194.5$ days.}
\label{figCompNimastep}
\end{figure*}

Then, given that both orbits obtained with our symplectic scheme and {\sc NIMASTEP} are very close to each other, we have decided to show the absolute difference between each Keplerian element and for each time step. Results can be seen in Fig. \ref{figCompNimastep}. For each time step and orbital element, the absolute difference has been computed each day. Both softwares having been developed using different units of time and distance, residual errors could have been introduced artificially. Anyway, we are pretty confident that our comparison does not suffer too much from this kind of numerical errors. From Fig. \ref{figCompNimastep}, it can be seen that absolute errors are quite small for the eccentricity, inclination, longitude of ascending node and argument of pericenter, no matter which time step is considered. The situation is slighlty different for the semi-major axis. In this case, the error is clearly bigger for larger time steps and increases linearly with time. This can be explained by the fact that {\sc NIMASTEP} is a non-symplectic integration scheme and that the energy is not fully preserved. Hence, while both integrators turn out to be really efficient on short time scales, our symplectic integrator does not suffer from a slight drift on the semi-major axis on long time scales. Nevertheless, let us remark that {\sc NIMASTEP} with time steps equal to $432$ and $864$~s yields excellent results, the drift on the semi-major axis being reasonably low after $190$ years ($8.4443 \times 10^{-2}$ km and $2.9404\times 10^{-1}$ km respectively with times steps $432$ and $864$~s).
One of the main advantages of our integrator is that it lets us obtain accurate results even with big time steps. As a comparison point, it turns out that the {\sc NIMASTEP} software cannot be used with time steps larger than $1160$~s for this particular set of initial conditions and perturbations.
       
\section{Earth's shadow modelling}
\label{secShad}

\begin{figure*}
\begin{center}
\includegraphics*[width=12cm]{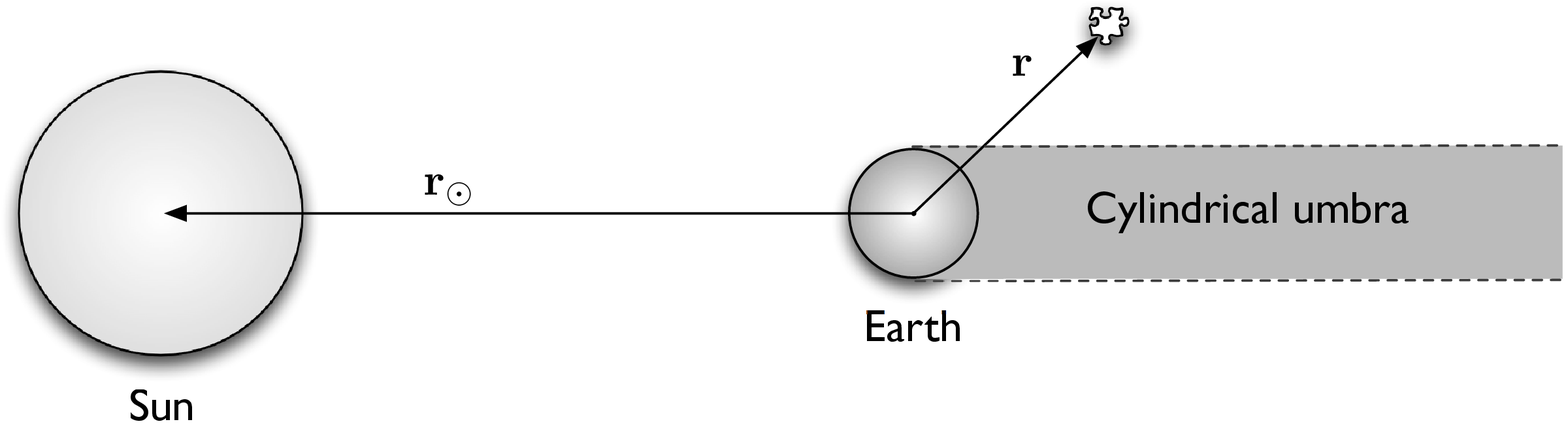} \\
\includegraphics*[width=12cm]{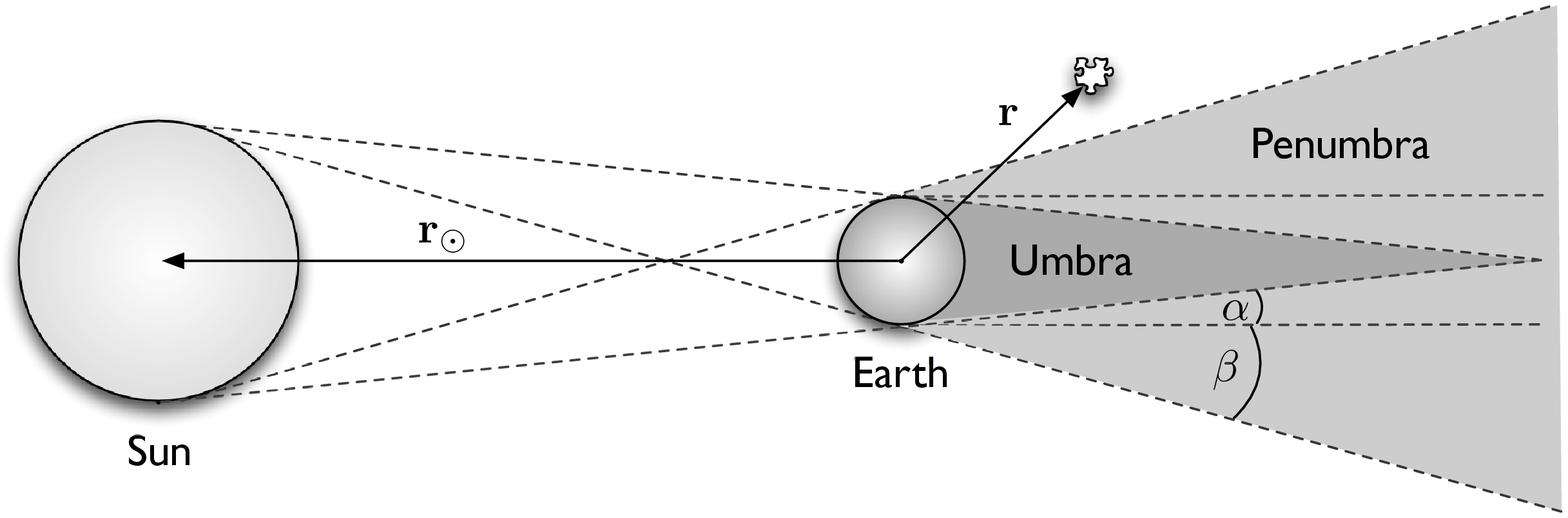} 
\end{center}
\caption{Top panel: cylindrical Earth shadow with solar rays assumed to be parallel when reaching the Earth. Bottom panel: Umbra-penumbra model including partial eclipses. Angles $\alpha$ and $\beta$ give the geometric difference between the cylindrical and the conical models.}
\label{CylConShad}
\end{figure*}

The Earth's shadow can first be modeled as a simple cylinder. In this case, the Sun is assumed to be far enough from the Earth and the solar rays are then supposed to be parallel. The geometry of this problem is illustrated in Fig. \ref{CylConShad} (top). However, a more realistic model includes the penumbra transition which lets us model partial eclipses (see Fig. \ref{CylConShad}, bottom). The distance to the Sun and diameters of both Earth and Sun have to be considered to compute the amount of sunlight actually reaching the space debris. In the following, we describe an existing method to model cylindrical shadows and present our solution (Section \ref{SecCyl}). Then, the same work is done in the case of penumbra transition models (Section \ref{SecPen}). Finally, in Sec. \ref{secComp}, we compare our model to the solution of \citet{Montenbruck2005} and we present an analysis of differences between orbits obtained with both our cylindrical and conical shadow models.

\subsection{Cylindrical shadow models}
\label{SecCyl}

In \citet{Escobal1976}, a method is proposed to find the orbital entrance and exit of a satellite from the shadow of the Earth. More precisely, two non-spurious roots of a quartic polynom in the cosine of the true anomaly correspond to the shadow entrance and exit (more details are given in \ref{appEscobal}). Then the effect of the solar radiation pressure can be switched on and off, depending on the angular position of debris on its orbit. Unfortunately, it turns out that this shadow crossing model can not be put directly in the equations of motion, yielding numerical errors on short time scales.

In order to avoid numerical errors in the integration process, it is necessary to find a smooth function $\nu({\mathbf r})$ equal to one when debris are in direct sunlight and zero otherwise. Then, ${\bf \nabla}_{\bf r} \mathcal H_{\text{srp}}({\mathbf r})$ can be multiplied by $\nu({\mathbf r})$ in equations of motion so that each shadow crossing is taken into account.

\begin{figure}
\begin{center}
\includegraphics*[width=9cm]{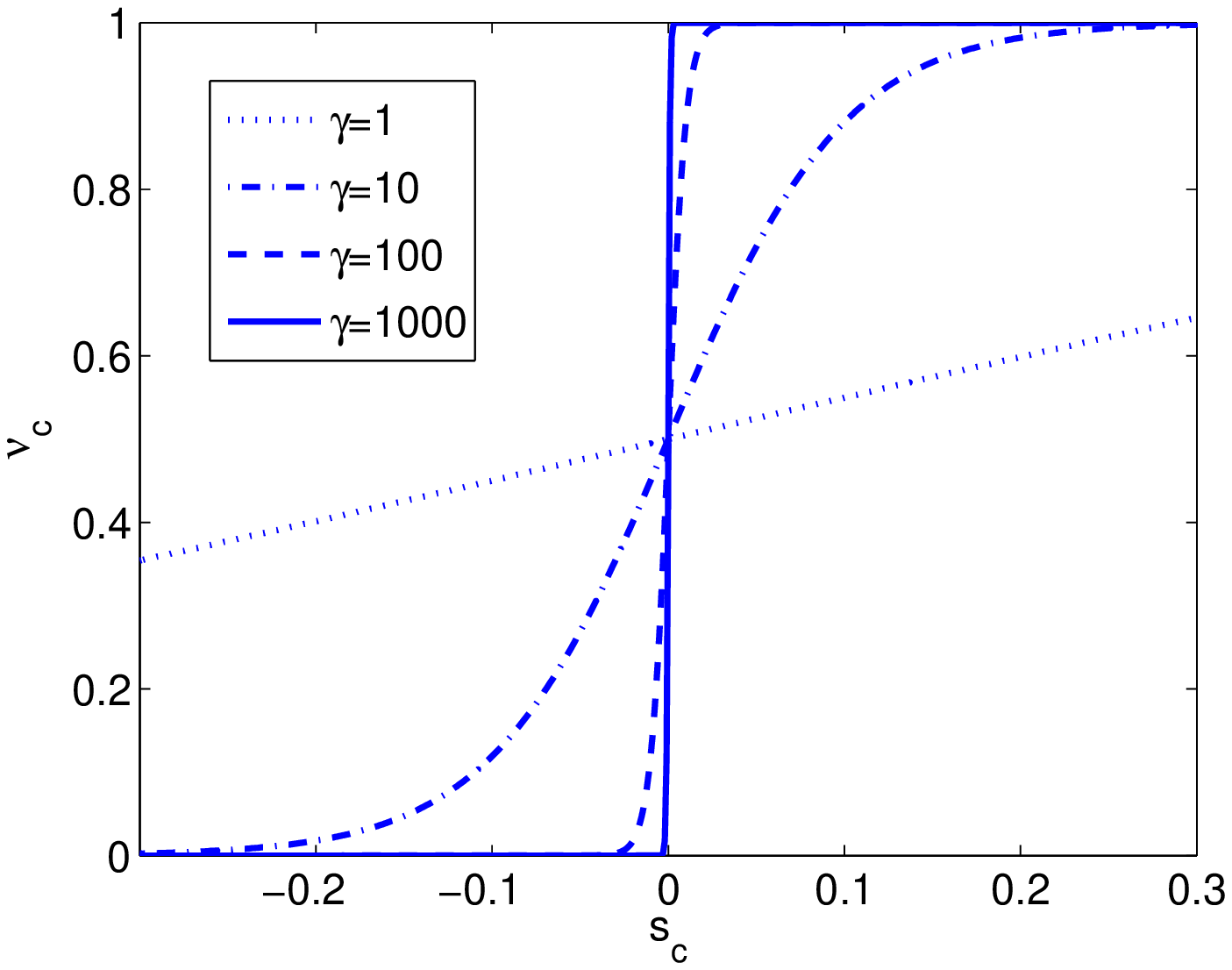} \\
\includegraphics*[width=9cm]{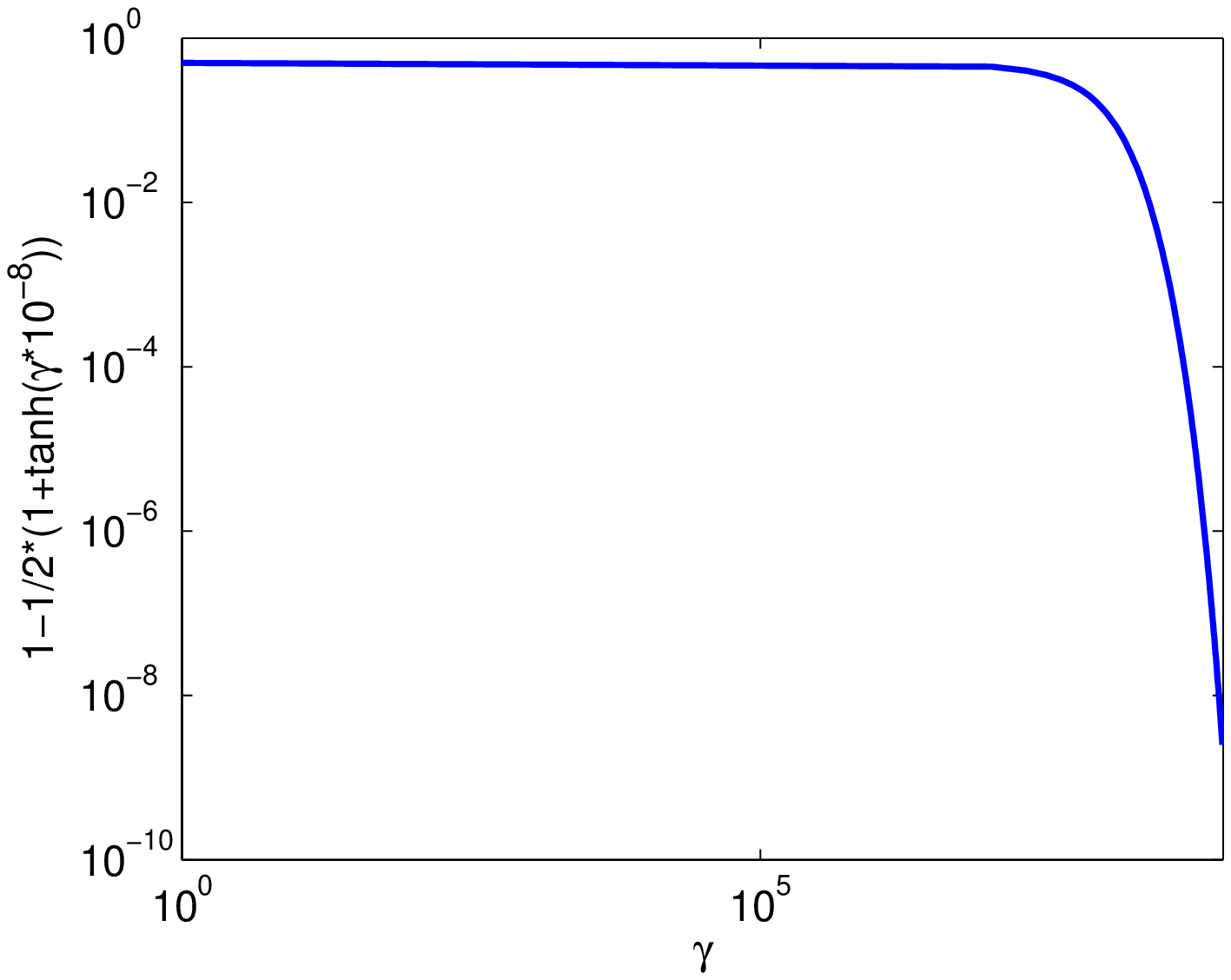}
\end{center}
\caption{Top panel : evolution of the function $\nu_{\text{c}}$ for different values of $s_{\text{c}}$ and of the parameter $\gamma$. Bottom panel : absolute difference between one and the function $\nu_{\text{c}}$ evaluated at $s_{\text{c}}(\mathbf{r})=10^{-8}$ for different values of the parameter $\gamma$.}
\label{tanhShadow}
\end{figure}

Starting from a preliminary relation in the model of \citet{Escobal1976}, it turns out that the satellite is situated in the cylindrical shadow of the Earth when
\begin{equation}
 s_{\text{c}}(\mathbf{r}):=\dfrac{\mathbf{r} \cdot \mathbf{r}_\odot }{r_\odot} + \sqrt{r^2-R_\oplus^2} \leq 0.  \label{eqsc}
\end{equation}
We introduce a new shadow function defined as
\begin{equation*}
 \nu_{\text{c}}({\mathbf r})=\dfrac{1}{2}\Big\{ 1+\tanh[\gamma \, s_{\text{c}}(\mathbf{r})]\Big\}= \left\{ \begin{array}{cl} 1 & \text{in cylindrical umbra} \\ 0 & \text{otherwise} \end{array} \right. 
\end{equation*}
where the constant $\gamma$ has to be fixed according to the required precision. The shape of this function is shown for different values of $\gamma$ in Fig. \ref{tanhShadow} (top). It can be seen that, the bigger $\gamma$, the sharper the function $\nu_{\text{c}}$. As a matter of fact, a perfect cylindrical shadow model would require $\gamma$ to be infinite. However, considering double precision floating point standard, Fig. \ref{tanhShadow} (bottom) shows that taking $\gamma=10^9$ is sufficient to represent cylindrical shadow crossings. Indeed, the absolute difference between $1$ and the function $\nu_{\text{c}}$ with $s_{\text{c}}(\mathbf{r})=10^{-8}$ is already of order $10^{-9}$.

As explained above, replacing ${\bf \nabla}_{\bf r} \mathcal H_{\text{srp}}({\mathbf r})$ by  $\nu_{\text{c}}({\mathbf r}) {\bf \nabla}_{\bf r} \mathcal H_{\text{srp}}({\mathbf r})$ in equations of motion turns out to be an efficient way to use the fully symplectic integration scheme and consider cylindrical-shaped shadows of the Earth. In this case, a special attention has to be paid to the integration time step. The latter must be small enough to perform some steps inside the umbra zone, which only represents a small part of the total revolution time. As illustrated in Fig. \ref{figComp}, the cylindrical shadow on a geostationary orbit only lasts around half an hour.
While this drawback cannot be neglected, it also lets us to use low order symplectic integrators. In particular, the $SBAB_2$ integrator is used when the shadowing effects are enabled, still keeping highly accurate results.

\subsection{Conical shadow models}
\label{SecPen}

Several attempts have been made to model the penumbra transition. A solution is proposed in \citet{Escobal1976} to add umbra-penumbra corrections. Basically, one ends up with a more accurate but also trickier shadow function which still induces numerical errors in our symplectic integration scheme because the radiation pressure does not include Earth's shadow as a smooth function.

Another kind of shadow crossing model can be found in \citet{Montenbruck2005}. In this case, a coefficient $\nu_M$ corresponds to the fraction of sunlight reaching the debris, based on the angular separation and diameters of the Sun and the Earth. Hence, $\nu_M$ is equal respectively to zero and one when the debris is in direct sunlight and umbra and corresponds to the remaining fraction of sunlight in the penumbra transition phase. It has to be noted that this fraction can only be defined in the penumbra cone. Hence, it is not possible to handle the function $\nu_M$ with a single formula at each time and it cannot be used directly within our symplectic scheme. Moreover, any stability study requiring the computation of the deviation vectors could not be used with this method, $\nu_M$ being a piecewise-defined function. Nevertheless, $\nu_M$ is kept back as a comparison criterion for our further developments.

\begin{figure}
\begin{center}
\includegraphics*[width=9cm]{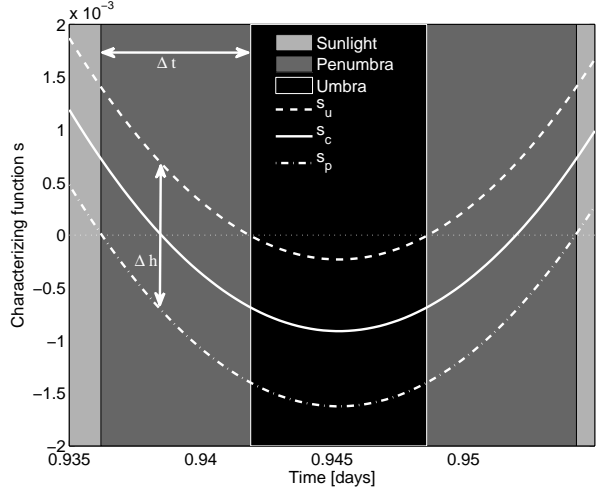}
\end{center}
\caption{Evolution of the functions $s_{\text{c}}$, $s_{\text{u}}$ and $s_{\text{p}}$ during a shadow crossing on a geostationary orbit. The penumbra cone is crossed when $s_p$ is negative and the space debris lays in the umbra cone while $s_u$ is negative. The time spent in the penumbra transition is noted $\Delta t$ and the difference between $s_\text{u}$ and $s_{\text{p}}$ at the entrance of the cylindrical shadow is denoted by $\Delta h$.}
\label{shadFunc}
\end{figure}

In the following, we present an innovative way of modelling umbra and penumbra cones crossings during the numerical integration of space debris orbit. First, we introduce angles $\alpha$ and $\beta$ representing the difference between the umbra cylinder and respectively the umbra and penumbra cones (see Fig. \ref{CylConShad})
\begin{eqnarray*}
 \alpha = \text{atan} \, \dfrac{R_\odot-R_\oplus}{\|\mathbf{r}-\mathbf{r}_\odot\|} \quad \text{and} \quad \beta = \text{atan} \, \dfrac{R_\odot+R_\oplus}{\|\mathbf{r}-\mathbf{r}_\odot\|}
\end{eqnarray*}  
with $R_\odot$ the radius of the Sun. Extending relation (\ref{eqsc}), it follows that space debris are in the umbra cone when
\begin{equation*}
 s_{\text{u}}(\mathbf{r}):=\dfrac{\mathbf{r} \cdot \mathbf{r}_\odot }{r_\odot} + \cos \alpha \left[\sqrt{r^2-R_\oplus^2\cos^2 \alpha}+R_\oplus \sin \alpha\right] \leq 0  \label{eqsu}
\end{equation*}
and in the penumbra cone when
\begin{equation*}
 s_{\text{p}}(\mathbf{r}):=\dfrac{\mathbf{r} \cdot \mathbf{r}_\odot }{r_\odot} + \cos \beta \left[\sqrt{r^2-R_\oplus^2\cos^2 \beta}-R_\oplus \sin \beta\right] \leq 0.  \label{eqsp}
\end{equation*}
An example of the evolution of functions $s_{\text{c}}$, $s_{\text{u}}$ and $s_{\text{p}}$ depending on time is shown in Fig. \ref{shadFunc}.

Now, we will show that the function $\nu_{\text{c}}$ can be adapted to include the penumbra transition. Actually, the parameter $\gamma$ will not be constant anymore but will be chosen so that the new shadow function $\nu_{\text{p}}$ is equal to one in direct sunlight, starts to decrease in the penumbra cone and is equal to zero in the umbra cone. The value of $1-\nu_{\text{p}}$ when the penumbra cone is crossed has to be fixed to attain a given precision treshold, denoted $\sigma$. Hence, we define the constant 
\begin{equation*}
 \delta:=\text{atanh}\,(1-\sigma).
\end{equation*} 
In this paper, $\delta$ is set equal to $8$, meaning that the precision treshold $\sigma\simeq 2.25 \times 10^{-7}$.

Then, assuming that the time spent in the penumbra transition, $\Delta t$, is known, $\gamma$ is set equal to $\delta/\Delta t$. With this configuration, 
\begin{equation}
 \nu_{\text{p}}({\mathbf r}) := \dfrac{1}{2}\left\{1+\tanh\left[ \dfrac{\delta}{\Delta t} s_{\text{c}}(\mathbf{r})\right]\right\} \label{eqnup}
\end{equation}
is smooth and such that
\begin{equation*}
  \left\{
  \begin{array}{rl} 
    \nu_{\text{p}}({\mathbf r}) = 1 & \text{if} \quad s_{\text{p}}(\mathbf{r}) > 0 \\
    \nu_{\text{p}}({\mathbf r}) = 1-\sigma & \text{if} \quad s_{\text{p}}(\mathbf{r}) = 0 \\
    \sigma \leq \nu_{\text{p}}({\mathbf r}) \leq 1-\sigma & \text{if} \quad s_{\text{p}}(\mathbf{r}) \leq 0 \quad \text{and} \quad s_{\text{u}}(\mathbf{r}) > 0 \\
    \nu_{\text{p}}({\mathbf r}) = \sigma & \text{if} \quad s_{\text{u}}(\mathbf{r}) = 0 \\
    \nu_{\text{p}}({\mathbf r}) = 0 & \text{if} \quad s_{\text{u}}(\mathbf{r}) \leq 0 
  \end{array}
  \right. .
\end{equation*}

The main difficulty lies in the way to estimate $\Delta t$. As a matter of fact, this quantity cannot be computed explicitely before each shadow crossing. However, we will show that it can be replaced by a quantity depending only on the position of the space debris. Both entrance and exit times spent in the penumbra cone being computed exactly in the same fashion, we will only explain our method in the entrance case. 

In following developments, each function $s_c$, $s_u$ and $s_p$ will be expressed as functions of the angle $\phi$ between ${\mathbf r}$ and ${\mathbf r}_\odot$. As a first step, we assume that $r$ does not depend on $\phi$, meaning that the orbit of the space debris is circular. It yields
\begin{eqnarray}
\hspace*{-1.1cm} s_\text{c}(\phi) & = & r \cos \phi + \sqrt{r^2-R_\oplus^2}  \\
\hspace*{-1.1cm} s_\text{u}(\phi) & = & r \cos \phi + \cos \alpha \left[\sqrt{r^2-R_\oplus^2\cos^2 \alpha}+R_\oplus \sin\alpha \right] \label{eqsuphi}\\
\hspace*{-1.1cm} s_\text{p}(\phi) & = & r \cos \phi + \cos \beta \left[\sqrt{r^2-R_\oplus^2\cos^2 \beta}-R_\oplus \sin\beta \right]. \label{eqspphi}
\end{eqnarray}
Let us also define $\phi_1$, $\phi_2$ and $\phi_3$ respectively as the value of $\phi$ at the entrance the cylindrical shadow, umbra and penumbra cones. Hence, the following relations hold: 
\begin{equation}
 s_\text{c}(\phi_1)=s_\text{u}(\phi_2)=s_\text{p}(\phi_3)=0. \label{relPhi}
\end{equation}
The difference ($ \Delta \phi$) between  $\phi_2$ and $\phi_3$ will help us to characterize $\Delta t$.
Technical details about the computation of $ \Delta \phi$ are given in \ref{appDeltaPhi}. It turns out that it can be approximated by  
\begin{equation}
 \Delta \phi := \phi_2-\phi_3 \simeq 2 \rho \dfrac{R_\odot }{r}.
\end{equation}
where $\rho:=r/r_\odot$.
Moreover, the link between $\Delta \phi$ and $\Delta t$ can be expressed as
\begin{equation}
 \Delta \phi = \dot \phi \Delta t \simeq \dfrac{2 \pi}{\text{day}} \Delta t. \label{eqDeltaPhi}
\end{equation}
Hence, 
\begin{equation}
 \Delta t \simeq \dfrac{\Delta \phi}{2 \pi} \, \text{day}\simeq\dfrac{\rho R_\odot}{\pi r} \, \text{day}.
\end{equation}
The final step is to show that the difference between $s_\text{u}$ and $s_{\text{p}}$ at the entrance of the cylindrical shadow, denoted by $\Delta h$, can be used instead of $\Delta t$. Simple calculations let us write down $\Delta h$ as
\begin{equation}
 \Delta h = 2 \rho R_\oplus \dfrac{R_\odot }{r} \label{eqDeltaH}
\end{equation}
Full details about this relation are to be found in \ref{appDeltaH}.

In conclusion, we have shown that $\Delta t$ can be approximated as
\begin{equation*}
  \Delta t \simeq \dfrac{\Delta h}{R_\oplus 2 \pi}.
\end{equation*}
From a geometrical point of view, our approximation means that the slope of each curve $s_{\text{u}}$, $s_{\text{c}}$ and $s_{\text{p}}$ is close to $-1$ at the shadow entrance and to $1$ at the shadow exit with our particular choice of units.

The same mathematical development can be achieved in the case of non-circular space debris orbits ($r$ depends on the angle $\phi$). The calculations are presented in \ref{appNonCircular}.

Going back to equation (\ref{eqnup}), it is easily seen that we now have
\begin{equation}
 \nu_{\text{p}}({\mathbf r}) = \dfrac{1}{2}\left\{1+\tanh\left[ \dfrac{\delta 2 \pi R_\oplus}{\Delta h(\mathbf{r})} s_{\text{c}}(\mathbf{r})\right]\right\} \label{eqnupmodif}
\end{equation}

Let us remark that, in practice, $\Delta h$ is computed with actual values of $\alpha$ and $\beta$ :
\begin{equation*}
  \begin{array}{rcl}
    \Delta h (\mathbf{r}) & = & s_\text{u}(\mathbf{r})- s_\text{p}(\mathbf{r}) \\
    & = & \cos \alpha(\mathbf{r}) \left[\sqrt{r^2-R_\oplus^2\cos^2 \alpha(\mathbf{r})}+R_\oplus \sin \alpha(\mathbf{r})\right]  \\
    &   & - \cos \beta(\mathbf{r}) \left[\sqrt{r^2-R_\oplus^2\cos^2 \beta(\mathbf{r})}-R_\oplus \sin \beta(\mathbf{r})\right].
  \end{array}
\end{equation*}
Also note that, with $\alpha=\beta=0$, $\nu_{\text{p}}$ is equal to $\nu_{\text{c}}$.

\subsection{Numerical comparisons}
\label{secComp}

\begin{figure}
\begin{center}
\includegraphics*[width=9cm]{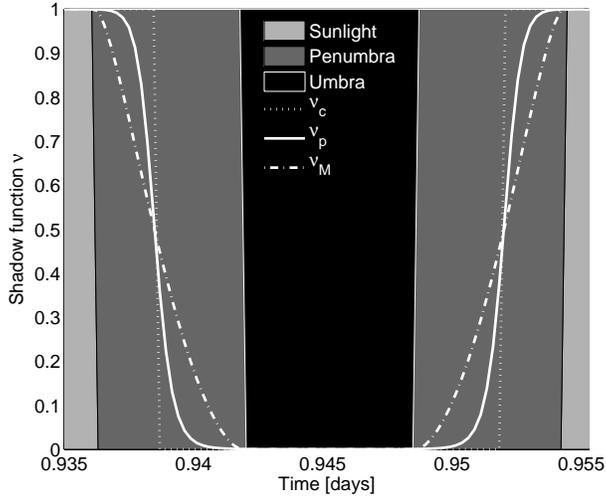}
\end{center}
\caption{Typical shadow crossing on a geostationary orbit. This figure shows the comparison between three shadow crossing models, as a function of time.}
\label{figComp}
\end{figure}

First, we compare our shadow functions to the one proposed in \citet{Montenbruck2005}. It is interesting to study the shape of these functions on a single shadow crossing. Considering a geostationary orbit, we show the evolution of $\nu_{\text{u}}$, $\nu_{\text{p}}$ and $\nu_{\text{M}}$ during a typical shadow crossing in Fig. \ref{figComp}. Even if the shapes of $\nu_{\text{p}}$ and $\nu_{\text{M}}$ are different, both share some common properties. It can be seen that both shadow functions cross the cylindrical shadow limit at $\nu_{\text{u}}=\nu_{\text{p}}=\nu_{\text{M}}\simeq 0.5$, start to decrease when entering the penumbra cone, are equal to zero in the umbra cone and increase again in the penumbra exit transition. A numerical comparison based on several shadow crossings is proposed later on.

\begin{figure*}
\begin{center}
\hspace*{-1cm}\includegraphics*[width=15cm]{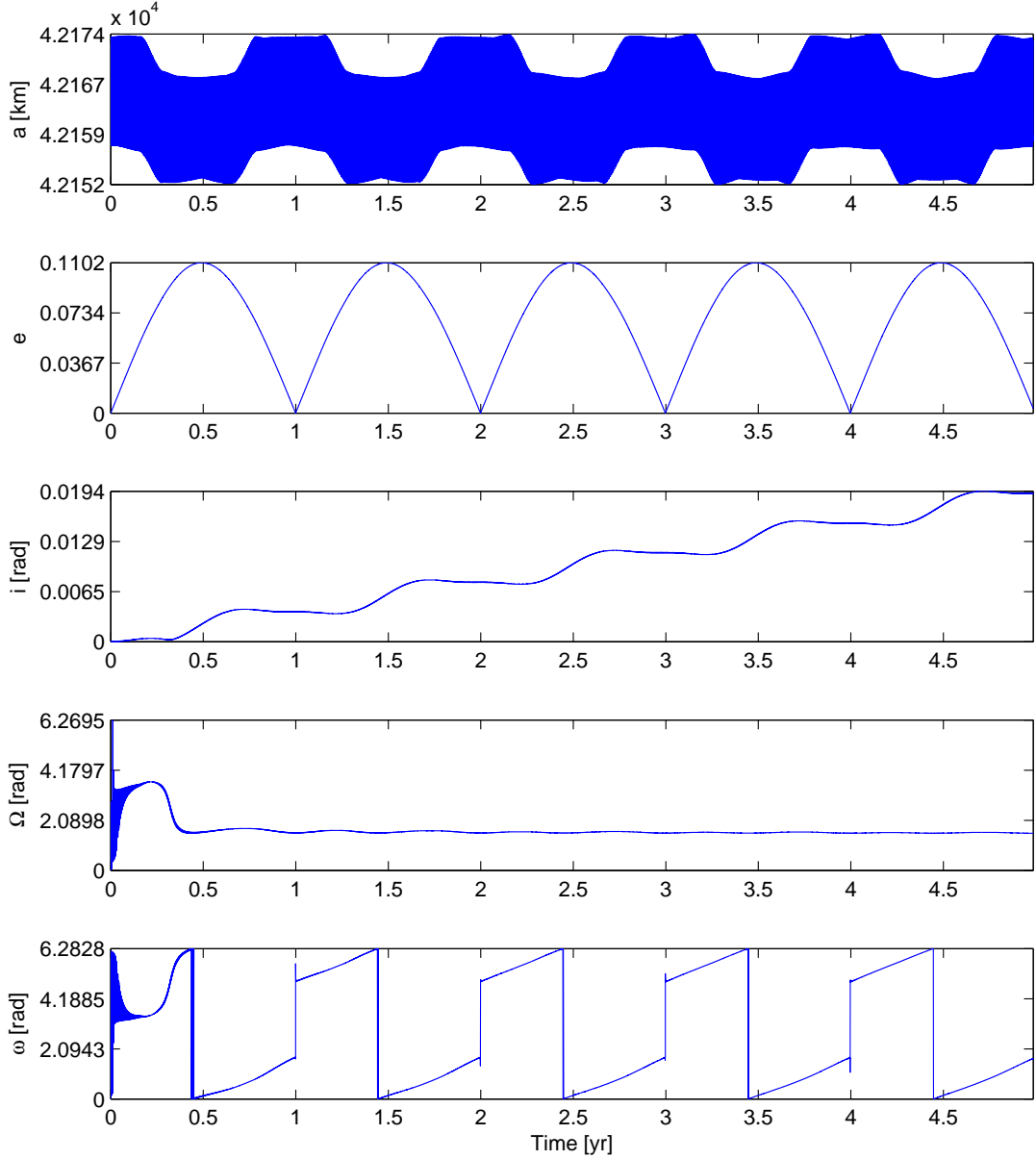}
\end{center}
\caption{Evolution of Keplerian elements of the orbit of space debris subject to the Earth's central attraction and solar radiation pressure. The initial semi-major axis is set at $42164$ km, the other elements are set equal to zero and the area-to-mass ratio is equal to $5$ m$^2$/kg. The shadow function $\nu_{\text{p}}$ is used to model the Earth's shadowing effects. The results are in agreement with what is proposed in Fig. 3 in \citet{ValkLemaitre2008}.}
\label{fig5yr}
\end{figure*}

Short-periodic effects of Earth's cylindrical shadows on the orbital elements of space debris have already been studied in \citet{ValkLemaitre2008}. For example, the evolution of the semi-major axis and eccentricity for space debris disturbed by the solar radiation pressure and an area-to-mass ratio equal to $5$ m$^2$/kg is described in \citet{ValkLemaitre2008} (Fig. 3). The same simulation has been performed in Fig \ref{fig5yr} with our conical shadow model. At first glance, it is not possible to detect significant discrepancies. 

\begin{figure*}
\begin{center}
\hspace*{-1cm}\includegraphics*[width=15cm]{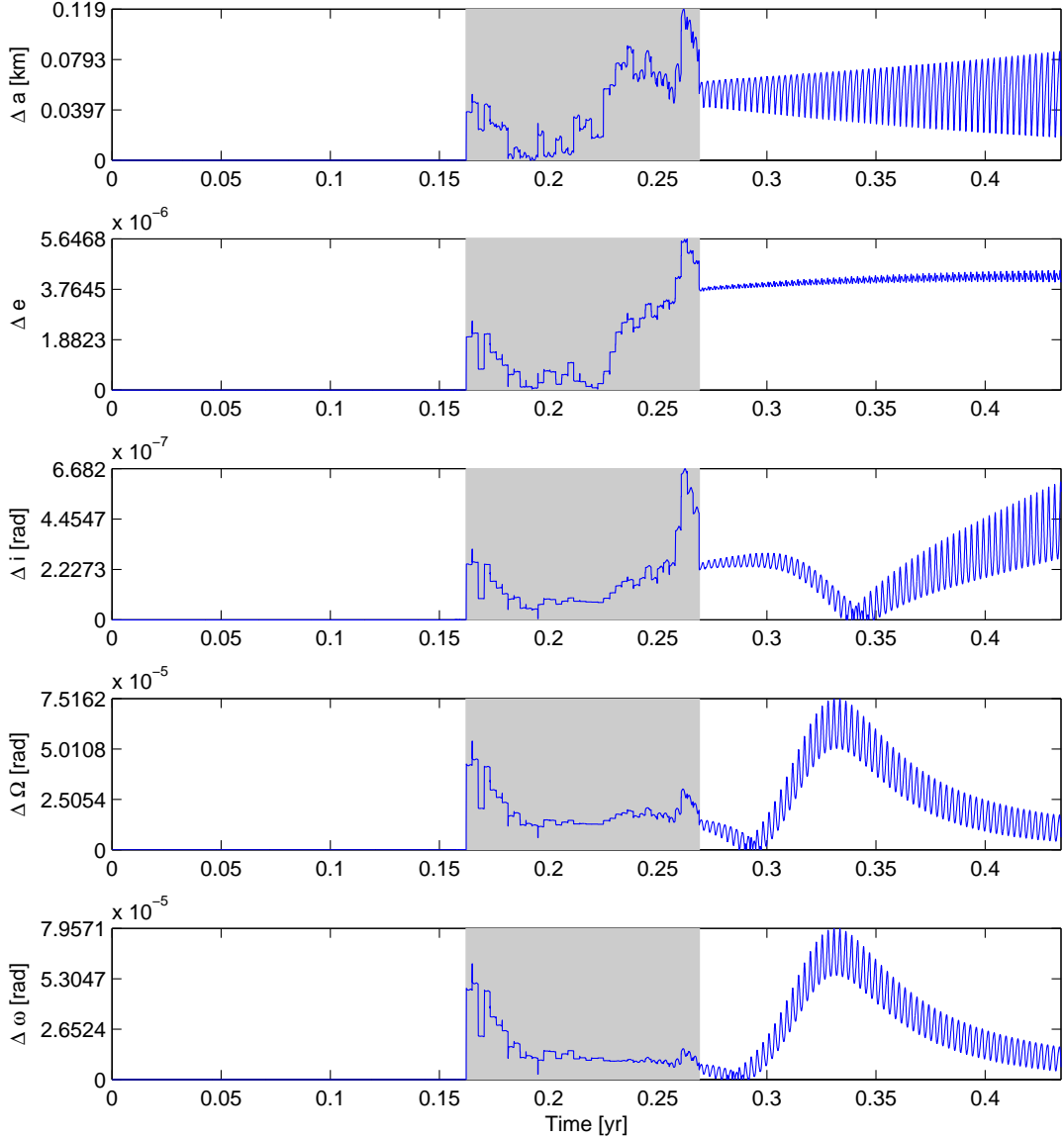}
\end{center}
\caption{Absolute difference between the orbital elements of space debris subject to the Earth's central attraction and solar radiation pressure with $\nu_{\text{c}}$ and $\nu_{\text{p}}$ shadow functions. The initial semi-major axes are $42164$ km, the other elements are set equal to zero and the area-to-mass ratio is equal to $20$ m$^2$/kg. This figure emphasizes the beginning of the difference between both orbits after the first shadow season represented by the gray zone.}
\label{figCompPenUm0.5yr}
\end{figure*}

\begin{figure*}
\begin{center}
\hspace*{-1cm}\includegraphics*[width=15cm]{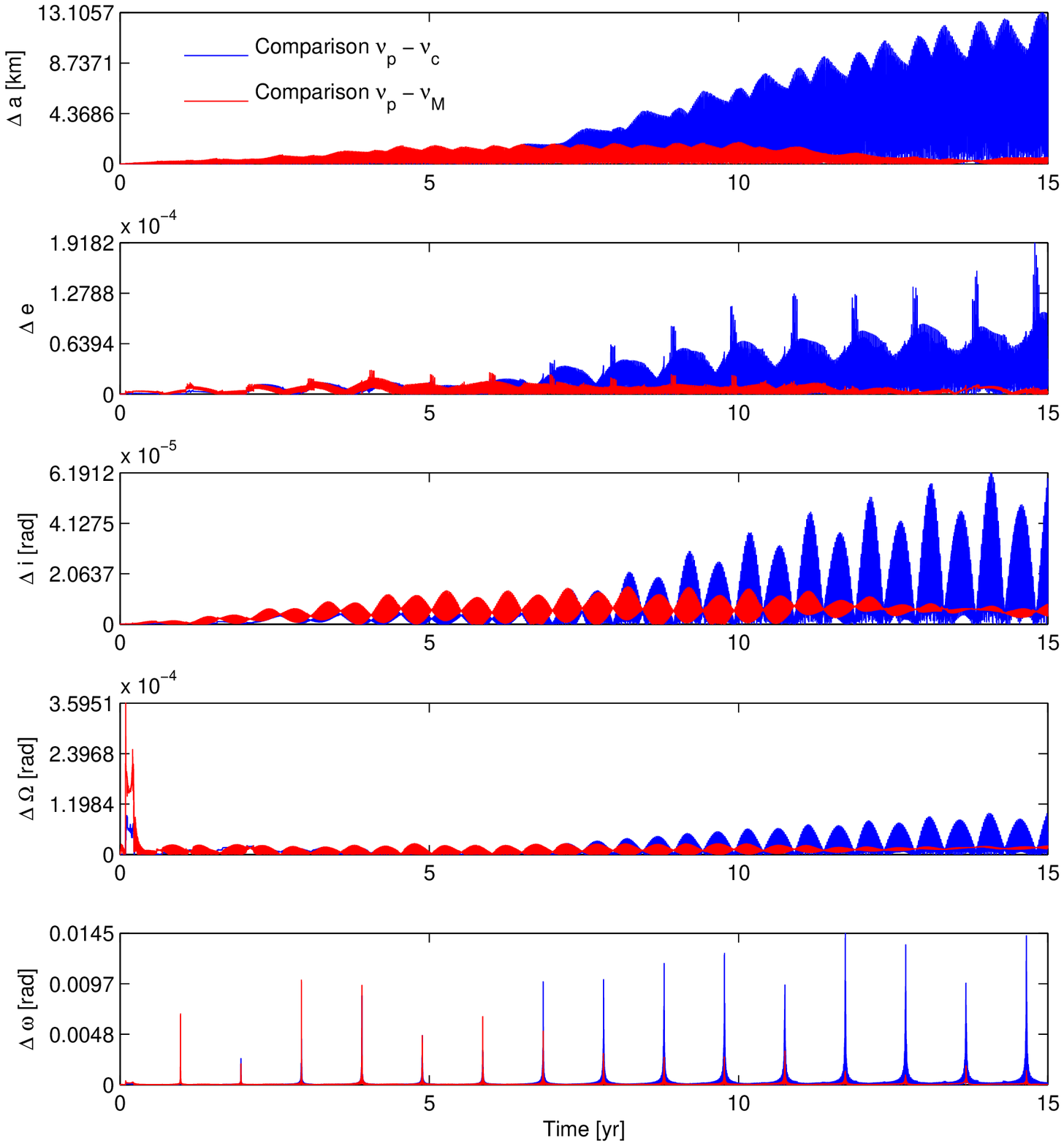}
\end{center}
\caption{Absolute difference between the orbital elements of space debris subject to the Earth's central attraction and solar radiation pressure with respectively $\nu_{\text{p}}$ and $\nu_{\text{c}}$ shadow functions (in blue) and $\nu_{\text{p}}$ and $\nu_{\text{M}}$ functions (in red). The orbit with $\nu_{\text{M}}$ has been computed by {\sc NIMASTEP} (ABM10). The initial semi-major axes are $42164$ km, the other elements are set equal to zero and the area-to-mass ratio is equal to $20$ m$^2$/kg. Each numerical integration has been performed with time steps equal to $150$ s.}
\label{figCompPenUm15yr}
\end{figure*}

Nevertheless, both shadow models clearly lead to significantly different debris trajectories. As illustrated in Fig. \ref{figCompPenUm0.5yr}, the absolute difference between orbital elements obtained with both models is zero before the first shadow season and starts to increase after this period of time. A shadow season appears each time that the Sun moves through the orbital plane of motion, leading to a succession of shadow crossings. More information about such phenomena can be found in \citet{ValkLemaitre2008}. The area-to-mass ratio is equal to $20$ m$^2$/kg for this example. After $15$ years, the difference between both trajectories keeps on increasing and reaches high values, especially for the semi-major axis (see Fig. \ref{figCompPenUm15yr}, blue curve). In light of this, it turns out that cylindrical shadow models are not reliable approximations of conical Earth's shadows, especially in the case of space debris with high area-to-mass ratios. The altitude of space debris trajectory must also be considered, the higher the orbit, the larger the time spent in the penumbra transition.

In Fig. \ref{figCompPenUm15yr}, the results obtained with both shadow functions $\nu_{\text{p}}$ and $\nu_{\text{M}}$ are represented by the red curve. Given that $\nu_{\text{M}}$ cannot be used with our symplectic scheme, it has been included in {\sc NIMASTEP} in order to perform the comparison between both methods. Differences between Keplerian elements in this case are clearly smaller than in the previous comparison involving $\nu_{\text{c}}$ and $\nu_{\text{p}}$ (Fig. \ref{figCompPenUm15yr}, blue curve). Moreover, this difference between both conical shadow models does not increase linearly with time. By way of conclusion, debris trajectories obtained with our symplectic integrator coupled to our smooth shadow function turn out to be consistent with the ones computed by {\sc NIMASTEP} (ABM10) using $\nu_{\text{M}}$.

\begin{figure}
\begin{center}
\includegraphics*[width=9cm]{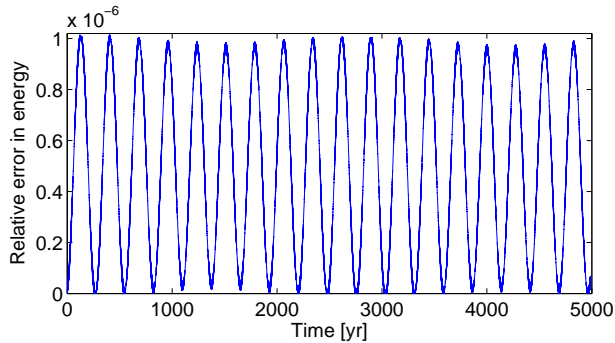}
\end{center}
\caption{Evolution of the relative error in energy of space debris motion subject to solar radiation pressure. The initial semi-major axis is set at $42164$ km, the other elements are set equal to zero and the area-to-mass ratio is equal to $0.1$ m$^2$/kg. The shadow function $\nu_{\text{p}}$ is used to model the Earth's shadowing effects.}
\label{figRelErrorEnergySrp}
\end{figure}

Eventually, a last remark is given about the energy conservation in the case of cylindrical and conical shadow models. The comments about the quasi-symplecticity are, of course, the same for the solar radiation pressure than for the third body contributions. Hence, we are still limited in the computation of the relative variation of the energy. However, it is shown in Fig \ref{figRelErrorEnergySrp} that the relative error in energy does not increase with time, even on an extremely long time span ($5000$ years). Let us remark that the energy computed takes into account the contribution of the solar radiation pressure with permanent sunlight. Indeed, the part of the Hamiltonian function corresponding to the solar radiation pressure cannot be retrieved from the equations of motion $\nu_{\text{u,p}}({\mathbf r})\,{\bf \nabla}_{\bf r} \mathcal H_{\text{srp}}({\mathbf r},\theta)$ , the latter being impossible to integrate analytically. A close look to the relative error in energy shows small perturbations during each shadow season but it does not result in a long term drift on the energy.

\section{Conclusion}

An efficient symplectic integration scheme has been presented in order to compute space debris motion. The underlying algorithm has been described and the accuracy of the integrator has been demonstrated by means of numerical comparisons. It has been pointed out that large time steps could be used and that the relative error in energy was not increasing with time, even on huge time spans. It should also be mentioned that our method is not stuck to one particular order and that it can adapted to the complexity of the perturbations and to the desired precision. Our integration scheme is able to take into account the Earth's gravitational potential, luni-solar and planetary gravitational perturbations and solar radiation pressure. In conclusion, it turns out that our algorithm represents a fast and reliable alternative to compute space debris or artificial satellites orbits, especially on long time scales.

We have also described an innovative method to model both cylindrical and conical Earth's shadow crossings by means of smooth shadow functions. We have explained why these ones were particulary convenient in the framework of symplectic integration. It has been shown that the cylindrical model was not a suitable approximation of conical shadows, especially in the case of space debris associated to high area-to-mass ratios. It has been noticed that one drawback of the computation of the shadow functions during the integration process relies in the obligation to reduce step sizes. Even if it can be considered as a limitation, one has to keep in mind that the order of the integrator can be reduced in order to speed up the calculations.

Future work will be devoted to stability studies involving shadowing effects. Such a task will be made easier by the fact that our shadow function derivatives are also smooth, enabling us to compute direclty the solutions of the variational equations. These relations are necessary to use chaos indicators like the MEGNO \citep{Cincotta03}. Hence we will be able to extend the work realized in \citet{Valk2009} by considering Earth's shadows.

\section*{Acknowledgements}
The work of Ch. Hubaux is supported by an FNRS PhD Fellowship. The authors would like to thank A. Rossi for helpful discussions and interactions. Numerical simulations were made on the local computing resources (Cluster iSCF) at the University of Namur (FUNDP, Belgium).

\appendix

\section{Orbital entrance and exit from the Earth's shadow}
\label{appEscobal}

In \citet{Escobal1976}, the entrance and exit true anomalies from the Earth's shadows are found to be the non-spurious solutions of the following function
\begin{equation}
 R_\oplus^2(1+e\cos f)^2+p^2(\beta \cos f + \xi \sin f)^2-p^2=0 \label{shadFuncSingular}
\end{equation}
where $f$ is the true anomaly, $p=a(1-e^2)$ is the \emph{semi-latus rectum} and $\beta$ and $\xi$ depend on the geocentric Cartesian position of the Sun and the inclination and longitude of the ascending node of the debris.
As explained in \citet{Escobal1976}, relation (\ref{shadFuncSingular}) corresponds to a quartic polynom in the cosine of the true anomaly. Hence, this function is transformed to standard form to find a new quartic polynom in $f$ which is solved in closed form by quadratic radicals (see Descartes' rule presented in \citet{Escobal1976}). 

The introduction of equinoctial elements $k_e=e \cos (\Omega+\omega)$ and $h_e=e \sin (\Omega+\omega)$ has been proposed in \citet{ValkLemaitre2008} to avoid singular orbital elements. It yields the following function
\begin{equation}
 R_\oplus^2(1+k_e\cos f+h_e \sin f)^2+p^2(\beta \cos f + \xi \sin f)^2-p^2=0 \label{shadFuncEqui}
\end{equation}
whose solutions are found using the so-called \emph{resultant method} (see \citealt{Gronchi2005} for an application example) . The latter lets us solve analytically the problem as a system of two algebraic equations in two variables.

We hereby propose a further improvement in the computation of the solutions of (\ref{shadFuncEqui}). By defining $T:=\tan(f/2)$, (\ref{shadFuncEqui}) can be written as
\begin{eqnarray*}
R_\oplus^2\left(1+k_e\dfrac{1-T^2}{1+T^2}+h_e\dfrac{2T}{1+T^2} \right)^2 \\ 
+p^2\left(\beta \dfrac{1-T^2}{1+T^2} + \xi \dfrac{2T}{1+T^2} \right)^2-p^2=0
\end{eqnarray*}
or, equivalently,
$$
\begin{array}{ccl}
  & &  T^4 [ p^2\beta^2-p^2+R_\oplus^2-2k_eR_\oplus^2+k_e^2R_\oplus^2 ]  \\
  & + & T^3 [ -4\beta\xi p^2+4h_eR_\oplus^2-4h_ek_eR_\oplus^2 ]  \\
  & + & T^2 [ 4\xi^2p^2-2p^2\beta^2-2p^2+4h_e^2R_\oplus^2+2R_\oplus^2-2k_e^2R_\oplus^2 ] \\
  & + & T [ 4\beta\xi p^2+4h_eR_\oplus^2+4h_ek_eR_\oplus^2 ] \\
  & + & [ p^2\beta^2-p^2+R_\oplus^2+2k_eR_\oplus^2+k_e^2R_\oplus^2 ] \\
  & = & 0
\end{array}
$$

Descartes' rule can then be used to find orbital entrance and exit true anomalies. The advantage of this method is twofold. First, the use of the tangent function directly indicates the right quadrant for the true anomaly. Then, the \emph{resultant method} is not necessary anymore.

\section{Computation of $\Delta \phi$}
\label{appDeltaPhi}

Relation (\ref{relPhi}) tells us that
\begin{equation}
  \cos \phi_2 + \cos \alpha \left[\sqrt{1-\dfrac{R_\oplus^2}{r^2}\cos^2 \alpha}+\dfrac{R_\oplus}{r} \sin\alpha \right] = 0 \label{cosphi2} 
\end{equation}
\begin{equation}
  \cos \phi_3 + \cos \beta \left[\sqrt{1-\dfrac{R_\oplus^2}{r^2}\cos^2 \beta}-\dfrac{R_\oplus}{r} \sin\beta \right] = 0. \label{cosphi3} 
\end{equation}
Then, we denote by $\rho$ the small quantity $r/r_\odot$. It leads to the following simplified expression
\begin{equation*}
 \dfrac{\|{\mathbf r}-{\mathbf r}_\odot\|^2}{r_\odot^2} = 1-2 \rho\cos\phi+\rho^2 \simeq 1-2 \rho \cos \phi 
\end{equation*}
where the small term $\rho^2$ has been neglected. Further calculations lead to the following expressions, keeping only terms of order $\rho$ :
\begin{equation}
 \sin \alpha \simeq  \tan \alpha \simeq \rho \, \dfrac{R_\odot-R_\oplus}{r} \quad \text{and} \quad \cos \alpha \simeq 1 \label{sinalpha}
\end{equation}
\begin{equation}
 \sin \beta \simeq  \tan \beta \simeq \rho \, \dfrac{R_\odot+R_\oplus}{r} \quad \text{and} \quad \cos \beta \simeq 1. \label{sinbeta}
\end{equation}
Replacing (\ref{sinalpha}) and (\ref{sinbeta}) values in (\ref{cosphi2}) and (\ref{cosphi3}), one obtains
\begin{eqnarray}
  \cos \phi_2 + \sqrt{1-\dfrac{R_\oplus^2}{r^2}}+ \rho R_\oplus\, \dfrac{R_\odot-R_\oplus}{r^2} & = & 0 \label{cosphi2simpl} \\
  \cos \phi_3 + \sqrt{1-\dfrac{R_\oplus^2}{r^2}}- \rho R_\oplus\, \dfrac{R_\odot+R_\oplus}{r^2} & = & 0. \label{cosphi3simpl}
\end{eqnarray}
By (\ref{cosphi2simpl}) and (\ref{cosphi3simpl}), we get
\begin{equation*}
 \cos \phi_3 - \cos \phi_2 = 2 \rho \, \dfrac{R_\oplus R_\odot }{r^2}.
\end{equation*}
Eventually, it can be shown that
\begin{equation}
 \Delta \phi := \phi_2-\phi_3 = 2 \rho \dfrac{R_\odot }{r}+\mathcal O (\rho (\phi-\phi_2)^2)
\end{equation}
where $\phi-\phi_2$ is small at the shadow entrance.

\section{Computation of $\Delta h$}
\label{appDeltaH}

At the cylindrical shadow entrance, the angle $\phi$ is equal to $\phi_1$. From (\ref{eqsuphi}) and (\ref{eqspphi}), one obtains
\begin{equation*}
 s_\text{u}(\phi_1) = r \cos \phi_1 + \cos \alpha \left[\sqrt{r^2-R_\oplus^2\cos^2 \alpha}+R_\oplus \sin\alpha \right] 
\end{equation*}
\begin{equation*}
  s_\text{p}(\phi_1) = r \cos \phi_1 + \cos \beta \left[\sqrt{r^2-R_\oplus^2\cos^2 \beta}-R_\oplus \sin\beta \right]. 
\end{equation*}
Then, from (\ref{sinalpha}) and (\ref{sinbeta}), it follows that
\begin{equation*}
 s_\text{u}(\phi_1) = r \cos \phi_1 + \sqrt{r^2-R_\oplus^2}+R_\oplus \rho \, \dfrac{R_\odot-R_\oplus}{r} + \mathcal O(\rho^2) 
\end{equation*}
\begin{equation*}
  s_\text{p}(\phi_1) = r \cos \phi_1 + \sqrt{r^2-R_\oplus^2}-R_\oplus \rho \, \dfrac{R_\odot+R_\oplus}{r} + \mathcal O(\rho^2). 
\end{equation*}
Eventually, (\ref{relPhi}) yields
\begin{equation*}
 s_\text{u}(\phi_1) = R_\oplus \rho \, \dfrac{R_\odot-R_\oplus}{r} + \mathcal O(\rho^2) 
\end{equation*}
\begin{equation*}
 s_\text{p}(\phi_1) = -R_\oplus \rho \, \dfrac{R_\odot+R_\oplus}{r} + \mathcal O(\rho^2). 
\end{equation*}
In conclusion, we can write
\begin{equation*}
 \Delta h = s_\text{u}(\phi_1)- s_\text{p}(\phi_1) \simeq 2 \rho R_\oplus \dfrac{R_\odot }{r}.
\end{equation*}

\section{Computation of $\Delta t$ in the non-circular case}
\label{appNonCircular}

The only difference with the circular case is given by the dependence of $r$ on $\phi$. Expressing $r$ in terms of Keplerian elements, one has
\begin{equation*}
 r(\phi) = \dfrac{a(1-e^2)}{1+e\cos f(\phi)}
\end{equation*}
where $f$ is the true anomaly. Expanding this relation (see e.g. \citealt{Murray1999}) and keeping only terms of first order in eccentricity, $r$ can be finally written as
\begin{equation*}
 r(\phi) \simeq a(1-e\cos M(\phi)) = a(1-e\cos (\phi+\psi_0))
\end{equation*}
where $M$ is the mean anomaly and $\psi_0$ is the appropriate phasing.

Following the same scheme as for the circular case, we end up with 
\begin{equation*}
 \Delta t = \dfrac{\rho R_\odot }{a\pi }(1+e\cos (\phi+\psi_0)) + \mathcal O(\rho^2 e+\rho e^2 + \rho^2 e^2) \quad \text{days}
\end{equation*}
and
\begin{equation*}
 \Delta h = \dfrac{2 \rho R_\oplus R_\odot }{a}(1+e\cos (\phi+\psi_0)) + \mathcal O(\rho^2 e+\rho e^2 + \rho^2 e^2).
\end{equation*}

\end{document}